\documentclass[preprint,12pt]{elsarticle}

\usepackage{amssymb}
\usepackage{amsmath}
\usepackage{graphicx}
\usepackage{color}
\DeclareGraphicsExtensions{.jpg,.png}

\journal{Commun. Nonlinear Sci. Numer. Simul.}

\begin{document}

\begin{frontmatter}



\title{Dynamic energy budget approach to evaluate antibiotic effects on biofilms}


\author{Bjorn Birnir}
\ead{birnir@math.ucsb.edu}
\address{Department of Mathematics, University of California, Santa Barbara, CA 93106, USA}
\author{Ana Carpio\fnref{fn1}}
\ead{carpio@mat.ucm.es}
\address{Departamento de Matem\'atica Aplicada, Universidad Complutense, 28040 Madrid, Spain, tel:+34-91-3944407, fax:+34-91-3944607}
\author{Elena Cebri\'an}
\ead{elenac@ubu.es}
\address{Departamento de Matem\'aticas y Computaci\'on, Universidad de Burgos, 09001 Burgos, Spain}
\author{Perfecto Vidal}
\ead{pervidal@ucm.es}
\address{Departamento de Matem\'atica Aplicada, Universidad Complutense, 28040 Madrid, Spain}

\fntext[fn1]{Corresponding author}

\begin{abstract}
Quantifying the action of antibiotics on  biofilms is essential to devise therapies against chronic infections. Biofilms are bacterial communities attached to moist surfaces, sheltered from external aggressions by a polymeric matrix. Coupling
a dynamic energy budget based description of cell metabolism to surrounding concentration fields, we are able to approximate survival curves measured for different antibiotics. We reproduce numerically stratified distributions of cell types within the biofilm and introduce ways to incorporate different resistance mechanisms. Qualitative predictions follow that
are in agreement with experimental observations, such as higher survival rates of cells close to the substratum when employing antibiotics targeting active cells or enhanced polymer production when antibiotics are administered. The current computational model enables validation and hypothesis testing when developing therapies.
\end{abstract}


\end{frontmatter}



\section{Introduction}
\label{sec:intro}

\begin{figure}
\centering
\includegraphics[width=8cm]{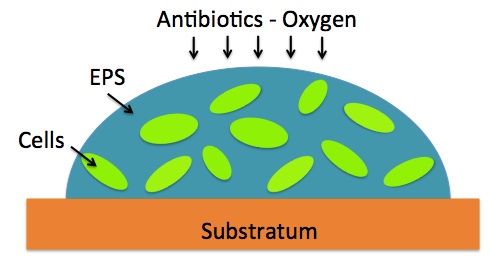}
\caption{{\bf Schematic representation of a biofilm}.
The film, formed by bacterial cells encased in a polymeric
matrix (EPS), adheres to a substratum and receives
nutrients, oxygen and antibiotics from the surrounding
flow.}
\label{fig1}
\end{figure}

Biofilms are bacterial aggregates that grow on moist surfaces, encased in a self-produced polymeric matrix, see Figure \ref{fig1}.
The matrix creates a favorable environment for their development,
facilitating nutrient, oxygen and waste transport \cite{matrix}.
It also acts as a shield against external aggressions by flows, disinfectants and antibiotics. The minimal bactericidal concentration  (MBC) and minimal inhibitory concentration (MIC) of antibiotics to  bacteria in their biofilm habitat may be up to 100-1000 fold higher compared with planktonic bacteria \cite{abbasmbic,hoibyresistance}.

Implant associated infections typically involve biofilm growth on the surface
of the implant \cite{vickerydevice}. They form on medical equipment
and prostheses, such as pacemakers and endotracheal tubes, central
lines, intravenous catheters, stents and artificial joints. Bloodstream infections, and many other hospital-acquired infections, may be caused by them. Biofilms may also spread on body surfaces such as heart valves (endocarditis), teeth, the lungs of cystic fibrosis patients (pneumonia), the middle ear and nose (otitis, rhinosinusitis), bones (osteomyelitis) or in chronic wounds \cite{hoibyresistance}. The biofilm matrix hinders phagocytosis and other actions of the inmune  system. Bacteria surviving standard antimicrobial  therapies are able to reproduce, originating chronic infections \cite{stoneburden}.
To tackle this problem, we must be able to understand how antibiotics
act and how resistance to antibiotics develops.

Antibiotics affect cells in diverse ways \cite{jaramayanantibioticaction}.
$\beta$-lactams (penicillins, cepha-losporins, carbapenems) and glycopeptides
(vancomycin) inhibit cell-wall synthesis.  Aminoglycosides (streptomycin,
gentamycin, tobramycin) inhibit protein synthesis. Quinolones (ciprofloxacin,
ofloxacin) inhibit DNA replication. Tetracyclines inhibit translation. 
Polymyxins such as colistin disrupt charge distributions in the outer cell 
membrane.
Eventually, the damage caused to the cell produces its death.
To exert their antibacterial action, antibiotics undergo a certain process.
They need to penetrate the cells,  remain stable and accumulate to reach
inhibitory concentrations.
Sometimes they have to take an active form.  After detecting their
target, they interact with it to exert their action. Interferences
in either of these processes may result in cell resistance to the
antibiotic. Resistance typically proceeds through efflux systems (the
antibiotic is pumped out of the cell), chemical alterations of the antibiotic
(cellular enzymes degrade it),  mutations in antibiotic target molecules,
and non-heritable resistance caused by environmental conditions \cite{daviesresistance,hoibyresistance,jaramayanantibioticaction,
stewartresistance3,stewartresistance4}. The main resistance
mechanisms for different types of antibiotics are summarized in
 \cite{jaramayanantibioticaction}.

The biofilm environment enhances bacterial resistance in a number
of ways. Biofilm development is influenced by quorum sensing  
\cite{daviesquorum}.
Through quorum sensing mechanisms,  bacteria sense when a critical
number of them are present in the environment. They respond by 
activating genes that produce exopolysaccharides \cite{kolter}. 
The polysaccharide matrix surrounding
the bacterial community delays diffusion of antibiotics inside the biofilm.
Nevertheless, direct measurements suggest that some antibiotics
equilibrate within the biofilm  \cite{broounresistance} after a waiting time.
{\em  Pseudomonas aeruginosa} tends to be the main source of
gram-negative infections in intensive care units in developed countries
\cite{surveillance}.
{\em P. Aeruginosa} and other bacteria express $\beta$-lactamase, an
enzyme that attacks $\beta$-lactams. An enzyme breaking the
antibiotic at a rate at which it crosses the cell membrane combined with delayed diffusion might explain resistance to penicillins, but not to other antibiotics \cite{broounresistance}.

As mentioned above, as we penetrate from the outer biofilm surface towards the interface with the substratum, gradients of oxygen and nutrients develop. Oxygen depletes \cite{debeeroxygen,werneroxygen}.
These gradients result in increased  doubling times for cell division
and reduced bacterial metabolic activity.  The intensity of metabolic
processes is stratified: high activity in the outer layers and slow growth
or no growth in the inner core. These dormant cells are partially
responsible for tolerance to antibiotics. Popular monotherapies with
$\beta$-lactams are only active against  dividing cells \cite{anwarenhanced},
forcing combinations with antibiotics that are active against nondividing
cells, such as colistin \cite{hoibyresistance}. Oxygen limitation and
metabolic rates are also important  factors enhancing the tolerance
of biofilms to ciprofloxacin and  aminoglycosides \cite{walterstolerance}.

In the biofilm, bacteria are exposed to oxidative stress, that causes
hipermutability. Enhanced production of reactive oxygen species
(ROS), either released in response to the infection or produced by the
alterations in the DNA repair system of the bacteria, leads to an
environment with low oxygen tension filled with oxygen radicals
\cite{hoibyresistance,mandsbergoxidative}. Augmented $\beta$-lactamase
synthesis, overexpression of efflux-pumps and increased EPS production
follow \cite{baggeincreasedeps,hoibyresistance,liinfluxeffluxpa}.
Studies with toxicants have also shown increased EPS production and ability
to adapt to the toxicant,  repairing damage to the cell \cite{halantolerance}.
Quorum sensing inhibitors \cite{hentzerinhibitquorum}, efflux inhibitors
\cite{lomovskayainhibitefflux,mahamoudinhibitefflux}, antioxidants
reducing the oxidative stress and mutations \cite{mandsbergoxidative},
together with enzymes able  to dissolve the biofilm matrix  \cite{alipourdissolve},
may provide strategies to overcome resistance mechanisms.

Mathematical modeling can assist in the design of therapies and the
interpretation of experimental data. Early models  were able to reproduce
elementary qualitative behavior. Reference \cite{stewartmodel} uses
coupled reaction-diffusion equations for the concentration of oxygen, antibiotics
and the volume fractions of live and dead cells to predict survival profiles
inside thick biofilms due to slow growth. Reference \cite{nicholsmodel}
predicts that the biofilm matrix can not prevent diffusion of $\beta$-lactam
antibiotics  into the bacteria provided the amount of chromosomal
$\beta$-lactamase is low. The diversity of the mechanisms involved in
biofilms resistance to antibiotics suggests the opportunity of adapting
dynamic energy budget (DEB) frameworks to describe the effect of antibiotics
on them. DEB models have already been exploited to describe
the effects of toxicant exposure on populations of floating bacteria
\cite{klanjscekplos}. Like antibiotics, toxicants interfere with the metabolism
of cells and increase the energy required for cell maintenance. The cell
requires additional energy to expel the toxicant and repair damage
caused by toxicant activity (DNA, RNA, protein repair).

Dynamic energy budget models relate biomolecular processes to individual
physiology and population dynamics \cite{kooijmanbook}.  They have
been successfully used to study scaling behaviors of all sorts of living beings, 
from plants and animals to cells \cite{birnirmyxo}.
Basic DEB models describe acquisition of  biomass and energy, as well as energy allocation for cell maintenance, growth and division. 
A standard description of aging and death processes is also available \cite{kooijmanbook}. Surplus reactive oxygen species (ROS) causes irreparable
damage to the cell. Damage components become `damage inducing compounds',
resulting in cell malfunctioning and further damage, raising the death probability.
Death and damage are represented through `hazard' and `aging acceleration'
variables. Antibiotic impact, like toxicant consequences, must be described
modifying standard fluxes and rates. To represent the effect of toxicants on
floating bacteria, Reference
\cite{klanjscekplos} included acclimation to the toxicant effect, environmental
degradation due to cell products, toxicant induced ROS production, and their
influence on assimilation, growth, aging and harzard rates. The only variable
states characterizing the environment are substrate and toxicant concentrations.

In this paper, we analyze antibiotic effects on {\em Pseudomonas aureginosa} 
biofilms by means of an adapted  DEB model. Unlike typical DEB
models for cells \cite{klanjscekplos}, we introduce here spatial variations in the concentration equations to study the local influence of the biofilm shape and composition. Spatial gradients in the concentrations of oxygen
and antibiotics are relevant to the bacterial survival profile in these
communities. We couple numerically the metabolism of individual cells
to the diffusion equations describing concentrations and 
some additional field. Our model accounts for EPS production,
identified as a cause of antibiotic resistance. The antibiotic acts on growth,
maintenance, aging and hazard rate in a similar way to toxicants. EPS
production increases acclimation, while reducing growth, damage and hazard rates, see Figure \ref{fig2}.


We have fitted the model for a quantitative agreement with the death rates 
reported for {\em P. aureginosa} under different antibiotics \cite{ishidadata}
and explored the spatial variations in the dead/alive cell distribution. Our
simulations reproduce expected qualitative trends. Augmenting the dose,
the number of dead cells increases. Increasing the EPS extent, the
number of dead cells decreases. Antibiotic presence enhances EPS
production. When we apply antibiotics targeting active cells a necrotic
region progressing from the outer biofilm layers to the inner biofilm
core appears. Instead, antibiotics targeting slowly growing cells would 
destroy first the inner dormant core.

The paper is organized as follows. Section \ref{sec:equations} describes
our spatially varying DEB model. Section \ref{sec:computational} presents
the computational setting.  Parameters are fitted to measured
death rates in Section \ref{sec:parameter}.
 Section \ref{sec:results}  discusses numerical simulations
illustrating the above mentioned qualitative behaviors. Finally, Section
\ref{sec:conclusions} states our conclusions.

\section{Basic DEB model for a biofilm including EPS production and antibiotics}
\label{sec:equations}

In this section we propose the basic equations for a dynamic energy budget 
theory (DEB) of a biofilm including polymer (EPS) production, coupled to diffusion 
of oxygen and antibiotic concentrations. We also introduce equations  describing the effect of the antibiotic on the bacteria. The framework is  similar to that developed in \cite{klanjscekplos} for the study 
of the effect of toxicants on a homogeneous population of floating bacteria.
Differently, in a biofilm we must incorporate EPS production, as well as 
diffusion processes in a number of spatially varying magnitudes, taking into
account the specific action mechanisms of antibiotics. Also, the population
of bacteria differentiates in several types. Only some of them become EPS
producers. The remaining bacteria grow and divide normally, unless resources 
are so scarce to trigger deactivation or damage is large enough
to kill them.

\subsection{The basic energies}

\begin{figure}
\centering
\includegraphics[width=10cm]{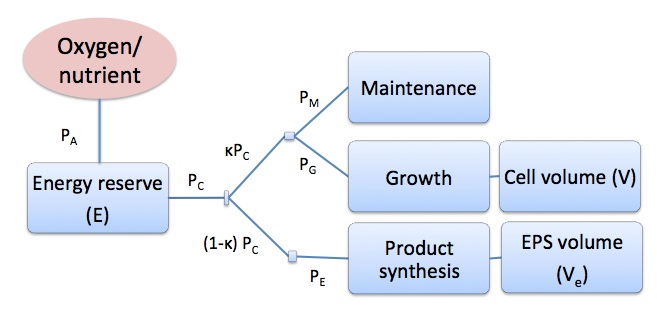}
\caption{{\bf The $\kappa$ rule for partition of energy in biofilm forming bacteria.} A fraction $\kappa$ of the total energy is devoted to maintenance
and growth of the cell. The remaining $1-\kappa$ fraction is invested in
polymer (EPS) synthesis.}
\label{fig2}
\end{figure}

DEB is essentially a scaling theory of different types of energies characterizing  the biofilm. The basic differential equations describe the fluxes of these  energies.  
We apply a variant of the $\kappa$ rule \cite{kooijmanbook}, see Figure \ref{fig2}. 
A $\kappa$ fraction of the energy is used for bacterial growth and division. Notice that bacteria do not mature and 
age, just grow and divide. A  $1-\kappa$ fraction of the energy is used for EPS production.

Each alive bacterium in the biofilm evolves according to the following system of equations, setting $\kappa=1$ for normal cells and $0 < \kappa < 1$ for EPS producers. 

\begin{enumerate}
\item {\it Scaled energy density  $e(t)$}: 
\begin{eqnarray}
{de \over dt} = \nu' (f-e),
\quad
f={C_o\over C_o+K_o}, \quad \nu'= \nu_{A} \nu,
\label{ec:scaledenergy1}
\end{eqnarray}
where $C_o$ is the oxygen limiting concentration, 
$K_o$ the limiting concentration half saturation value, 
$f$ is the scaled functional response, 
$\nu$ is the energy conductance 
and $\nu'$ takes into account toxic effects on conductance 
through $\nu_A$, 
defined in equation (\ref{ec:reducedconductance}).

\item {\it Dimensionless cell volume  $v(t)$}: 
\begin{eqnarray}
{dv \over dt} = (r {a\over a_M} - h)v, \quad
r= \left({\nu' e - m_{\kappa}  g_{\kappa}
\over e+g_{\kappa} }\right)^+,
\label{ec:scaledcellvolume1}
\end{eqnarray}
where $r$ is the bacterial (cell biomass) production rate, 
$m_{\kappa}$ is the maintenance rate 
and $g_{\kappa}$ is the investment ratio. 
The remaining parameters and magnitudes are linked to environmental toxicity, 
in our case antibiotic concentration, see below. 
$h$ is the hazard rate, 
$a$ 
is the acclimation energy density and  
$a_M$ is the target acclimation energy. 
The symbol $^+$ stands for positive part, which becomes
zero for negative expressions.

Bacteria are rod-shaped, and grow in length. If we fix an average radius, equation (\ref{ec:scaledcellvolume1}) provides an equation
for the time evolution of the cell length $l$.


\item {\it Dimensionless volume of EPS { $v_e(t)$}}: 
\begin{eqnarray}
{dv_e \over dt} =  {g_{\kappa}\over g_{e,\kappa}} (r {a\over a_M} - h) v
+ {m_{\kappa}\over g_{e,\kappa}} v = r_e v.
\label{ec:scaledepsvolume1}
\end{eqnarray}
This equation defines a rate of EPS production $r_e$.
In absence of  acclimation and hazard effects ($a=a_M$, $h=0$),
we set the reference rate $r_e=k r + k',$ that can be fitted to experiments,
as we explain in Section \ref{sec:results}.

\end{enumerate}
The dimensionless magnitudes $e,v$ and $v_e$  represent the fundamental energies characterizing the cell (bacterium) and its EPS production. Their fluxes are written here for each individual cell, $e$ being the total energy available for both growth and reproduction and for EPS production. $v$ keeps track of the volume of the cell as it grows, this is the energy available for reproduction by division. $v_e$ keeps track of the energy available to form the matrix that constitutes the biofilm and protects the bacterium.   
They may be related to dimensional volumes $V$ and $V_e$
multiplying by a characteristic cell volume $V_m$.
The dynamics of these variables is coupled to the evolution of other magnitudes, governed by equations we introduce next. Table \ref{table1} collects all the variables  appearing in the model. Table \ref{table2} gathers the parameters, specifying units and selected values. The discussion in Section \ref{sec:parameter}  suggests that a characteristic time scale for  
equations (\ref{ec:scaledenergy1})-(\ref{ec:scaledepsvolume1}) is set by the bacterial doubling time  $\sim {1 \over \mu_{max}}$ or the conductance
$\nu$, in a time scale of hours. Diffusion processes for oxygen and antibiotic concentrations, however, evolve in a time scale of seconds, as concluded from the equations presented next and the parameter values in Table \ref{table2}.

\begin{table}[h!] 
\centering
\begin{tabular}{|c|c|c|c|c|}
\hline 
Symbol & Units & Variable  &    \\
\hline 
t & s & time & \\
x  &  m &  space & \\
& & & \\
\hline 
Symbol & Units & State variable  & Equation  \\
\hline
{
$e$} &  n.d. & Scaled energy density  & Eq. (\ref{ec:scaledenergy1})  \\
$v$ & n.d. & Dimensionless bacterial volume & Eq. (\ref{ec:scaledcellvolume1})  \\
$l$ & n.d. & Dimensionless bacterial length & Eq. (\ref{ec:scaledcellvolume1})  \\
$v_e$ & n.d. & Dimensionless EPS volume & Eq. (\ref{ec:scaledepsvolume1}) \\
$\varepsilon$ & n.d. & Environmental degradation  & Eq. (\ref{ec:degradation}) \\
$a$ & n.d. & Acclimation energy density & Eq. (\ref{ec:acclimation}) \\
$h$ & ${\rm hr}^{-1}$ & Hazard rate & Eq. (\ref{ec:hazard}) \\
$q$ & ${\rm hr}^{-2}$ & Aging acceleration & Eq. (\ref{ec:aging}) \\
$C_{o} $ &  ${{\rm mg} \over \ell} $ & Oxygen concentration 
& Eq. (\ref{ec:oxygen}) \\
$C_{a}$ & ${{\rm mg} \over \ell}$ & Antibiotic concentration 
& Eq. (\ref{ec:antibioticbio}) \\
$ [C_{IN}] $  &  ${{\rm mg} \over {\rm Cmol}} $ & 
Antibiotic cellular density & Eq. (\ref{ec:antibiotic}) \\ 
& & & \\
\hline
Symbol & Units & Auxiliary variable  & Equation  \\
\hline
$ f $  &  n.d. &  Scaled functional response &   
Eq. (\ref{ec:scaledenergy1})   \\
$ r$  &  ${\rm hr}^{-1}$ & Bacterial production rate & 
Eq. (\ref{ec:scaledcellvolume1}) \\
$\nu'$  &  ${\rm hr}^{-1} $ & Conductance modified by exposure & 
Eq. (\ref{ec:reducedconductance}) \\
\hline
\end{tabular}
\caption{ {\bf Variables of the model.} The state variables
define the model. The auxiliary variables are introduced for simplicity.
Non-dimensional variables are labeled 'n.d.'. }
\label{table1}
\end{table}

\subsection{ Diffusion of oxygen and antibiotics}
\label{sec:oxygen}

Let  $C_o(\mathbf x, t)$  denote the concentration of oxygen and   $C_a(\mathbf x, t)$  the concentration  of antibiotic. 
These quantities are described by two diffusion equations.  
The antibiotic accumulated inside the cell is described using the antibiotic cellular density  $[C_{IN}](t)$, as discussed next.

\begin{enumerate}
\item {\it Oxygen concentration 
diffusion inside the biofilm}:
\begin{eqnarray}
 {\partial C_o \over \partial t} 
= d_o \Delta C_o - {\mu_{max} f \over Y_{x/o} } X 
- {(k \mu_{max} f+k') \over Y_{p/o}} X,
\label{ec:oxygen}
\end{eqnarray}
where $\mu_{max}$ is the maximum specific growth rate. 
$k$ is the growth associated polymer formation rate coefficient  
and $k'$ is the non-growth associated polymer formation rate coefficient. 
$Y_{x/o}$ is the cellular yield coefficient of $C_o$ 
and $Y_{p/o}$ is the polymer  yield coefficient of $C_o$. 
$d_o$  
is the oxygen diffusion coefficient inside the biofilm.
$X$ is the cellular structure concentration 
computed in a control volume $V_T$ containing $N$ cells:  
$X= \rho_x \sum_{i=1}^N V_i/V_T$, $\rho_x$ being the density of a cell.
When the volume control contains just one cell, we set $X= \rho_x v$, where
$v$ is the dimensionless cell volume. Equation (\ref{ec:oxygen}) is supplemented
with the boundary condition $C_o=C_{o,out}$ at the interface
with the oxygen providing fluid and the no-flux condition
${\partial C_o \over \partial \mathbf n}=0$ at the biofilm/substratum interface.

\item  {\it Antibiotic concentration 
diffusion inside the biofilm}:
\begin{eqnarray}
{\partial C_a\over \partial t}
= d_a \Delta C_a -  (R_e+R) C_a,
\label{ec:antibioticbio}
\end{eqnarray}
where $R_e$ and $R$ are evaluated averaging $r$ and $r_e$ in the control volumen $V_T$. $d_a$ is the antibiotic diffusion coefficient.
Equation (\ref{ec:antibioticbio}) is supplemented with boundary condition $C_a=C_{a,out}$ at the interface with the antibiotic providing fluid and 
the no-flux condition ${\partial C_a \over \partial \mathbf n}=0$
at the biofilm/substratum interface.

\item {\it Antibiotic cellular density}:
\begin{eqnarray}
{d[C_{IN}] \over dt} = k_A^I C_{a} - k_A^O [C_{IN}],
\label{ec:antibiotic}
\end{eqnarray}
where $k_A^I$ is the antibiotic influx coefficient, 
and $k_A^O$ is the antibiotic efflux coefficient. 
$C_a$ should be the antibiotic concentration outside the cell, that we take 
equal to $C_a$ given by (\ref{ec:antibioticbio}) in the control volume that 
 contains the cell. We may set $[C_{IN}(0)]=0$ initially.
\end{enumerate}

In view of the parameter values  in Table \ref{table2}, the chemical concentrations $C_o$ and $C_a$ will evolve much faster than the other magnitudes related to cell behavior, that is, in a time scale of seconds, not hours.

We have to complete the above equations by describing how the antibiotic and oxygen concentration act on the cell and influence its energy fluxes. We mostly follow \cite{klanjscekcadmium,klanjscekplos} considering the antibiotic to be a toxicant. Besides, we introduce several new DEB variables and make small modifications to account for spatial variations.

\subsection{The effect on cells of oxygen and antibiotics}
\label{sec:effect}

\begin{figure}[h!]
\centering
\includegraphics[width=11.5cm]{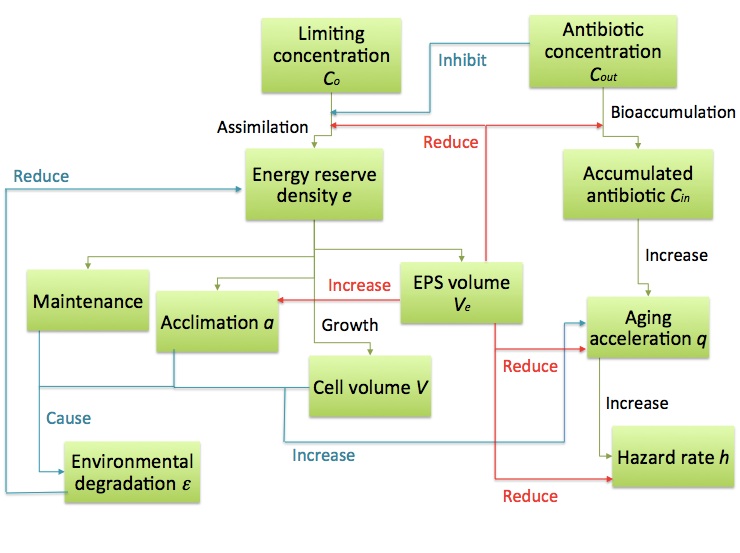}
\caption{{\bf Outline of the model.} Diagram summarizing the
main variables and their interactions.}
\label{fig3}
\end{figure}

The previous equations take into account the decline
in the organism's capacity to acquire and use energy due to
respiration and to the presence of antibiotics through a diminished
conductance, reduced growth due to acclimation to the antibiotic
and the hazard rate, see Figure \ref{fig3}.

\begin{enumerate}
\item{\it Conductance modified by exposure $\nu'$}: 
\begin{eqnarray}
\nu'=\nu_A\nu, \quad \nu_{A} = e^{-\gamma_{\varepsilon} \varepsilon}
\left(  1 + {C_a\over K_V} \right)^{-1},
\label{ec:reducedconductance}
\end{eqnarray}
where  $K_V$ is the noncompetitive inhibition coefficient
and $\gamma_{\varepsilon} $ is the environmental degradation effect
coefficient. 

\item {\it Environmental degradation  
$\varepsilon(\mathbf x, t)$}: 
\begin{eqnarray}
{\partial \varepsilon \over \partial t}
= d_{\varepsilon} \Delta \varepsilon +
\nu_{\varepsilon} (R+\nu_m m_{\kappa}) X.
\label{ec:degradation}
\end{eqnarray}
$\nu_{\varepsilon}$ is the environmental degradation coefficient
and $\nu_m$ is the maintenance respiratory coefficient. 
$X$ is the cellular structure concentration
computed in the control volume $V_T$ containing $N$ cells. $R$ is 
similarly computed averaging $r$ over this control volume. 
The diffusive term accounts
for the fact that $\varepsilon$ feels the spatial variations. We impose
no-flux boundary conditions.

\item {\it Acclimation energy density $a(t)$}: 
\begin{eqnarray}
{da \over dt} = (r+r_e) \left(1-{a \over a_M} \right)^+,
\label{ec:acclimation}
\end{eqnarray}
where $a_M$ is the target acclimation energy 
and $^+$ stands for the positive part.

\item {\it Hazard rate $h(t)$}: 
\begin{eqnarray}
{dh \over dt} = q - (r+r_e) h.
\label{ec:hazard}
\end{eqnarray}

\item{\it Aging acceleration $q(t)$}: 
\begin{eqnarray}
{dq\over dt}= e(s_q \,X\,q + h_a) (\nu'-r) + \left( {dq \over dt} \right)_{A} -rq,
\label{ec:aging}
\end{eqnarray}
where $h_a$ 
is the Weibull aging acceleration and $s_q$ is a 
multiplicative stress coefficient. 

\item{\it Aging in acclimation due to dissolved antibiotic and
EPS}: 
\begin{eqnarray}
\left( {dq \over dt} \right)_{A} =  k_{qA}^I [C_{IN}] -r_e q,
\label{ec:agingacclimation}
\end{eqnarray}
where $k_{qA}^I $ is the dissolved antibiotic toxicity. 
\end{enumerate}

\begin{figure}[h!]
\centering
\includegraphics[width=10cm]{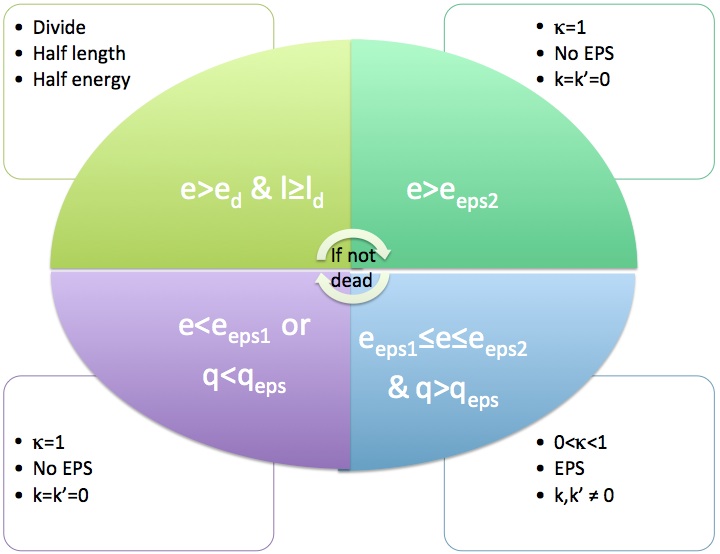}
\caption{{\bf The rules of cell division and differentiation.} Bacterial
cells act in different ways according to their energy, size and
aging rate.}
\label{fig4}
\end{figure}


\subsection{Cell division, differentiation and death}
\label{sec:division}

Bacteria of the {\em Pseudomonas Aeruginosa} genus are rod-like. Their volume
is defined by $V=\pi \rho^2 L$, $\rho$ being the radius and $L$ the length.
Increase in volume is equivalent to increase in length, setting $\rho$ equal to 
their average radius. Reference \cite{robinsonepsrates} reports cell lengths $L$ between $1.67$ $\mu$m and $2.75$ $\mu$m  for {\em P. Aeruginosa}. Diameters $2\rho$ vary between $0.5$-$0.73$ $\mu$m without clear relation to the length and with a preference for $2\rho=0.6$ $\mu$m. Notice that the equation for length is formally the same as the equation for volume (\ref{ec:scaledcellvolume1}), up to a factor. We may use $V_m=\pi \rho^2 L_m$, $L_m$ being the maximum length, to nondimensionalize.

In dimensionless variables, cell division happens provided a threshold  maintenance energy $e_d$ is surpassed when a threshold length $l_d$ is reached, see Figure \ref{fig4}. Then the cell divides and creates daughter cells.

Nondividing cells behave in different ways according to their lengths, energies, aging and hazard rates, see Figure \ref{fig4}:
\begin{itemize}
\item For intermediate energies $e_{eps1} \leq e \leq e_{eps2}$ the cells produce EPS according to equation (\ref{ec:scaledepsvolume1}), provided they are not newly born. A way to keep track of their age is the aging rate $q$, governed by (\ref{ec:aging}).
It is set initially equal to zero for newborn cells. We require $q \geq q_{eps}$ for the cell to have the ability to become a EPS producer.
\item For low energies $e \leq e_{eps1}$, the cell just maintains itself.
\item The hazard rate,  given by (\ref{ec:hazard}), determines when cells die.
Their survival probability  $p(t)$  
is governed by the equation
\begin{eqnarray}
{dp \over dt} = - hp, \quad p(0)=1.
\label{ec:survival}
\end{eqnarray}
\end{itemize}

Now with the energies $e,v,v_e$, oxygen and antibiotic concentrations $C_o$,$C_a$, $[C_{IN}]$ and the variables $\epsilon,a,h,q$, which measure how such concentrations influence the energies, we have a complete set  of variables and of equations. These additional rules define the behavior of each cell.

\section{Computational framework}
\label{sec:computational}

\begin{figure}[h!]
\centering
\includegraphics[width=12cm]{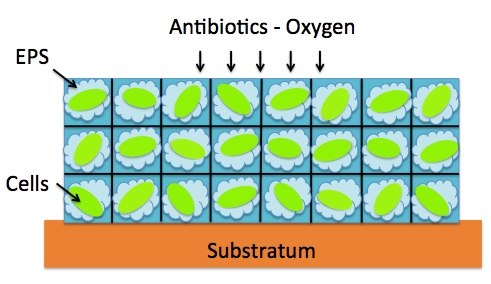}
\caption{{\bf Schematic description of the computational arrangement. }Continuous equations for concentrations are discretized in the square grid representing the biofilm.
Each tile contains cells of certain volume and energy, with an associated EPS volume around it, whose dynamics is governed by the DEB equations. Here, we assume that each tile contains just one cell. In practice, all these items are stuck together in a more amorphous mass. This is a simple computational idealization allowing us to combine macroscopic and microscopic information.}
\label{fig5}
\end{figure}

For computational purposes, the biofilm is identified with a slab containing cells. We divide the slab in a grid of cubic control volumes, that we use to discretize 
the concentration fields evaluating their spatial changes. A control volume $V_T$ may contain several cells, or just one, see Figure \ref{fig5}.

Variables $e$, $v$, $v_e$, $[C_{IN}]$, $a$, $h$, $q$, $p$ are assigned to each cell, governed by equations (\ref{ec:scaledenergy1}), (\ref{ec:scaledcellvolume1}), 
(\ref{ec:scaledepsvolume1}), (\ref{ec:antibiotic}), (\ref{ec:acclimation}),
(\ref{ec:hazard}), (\ref{ec:aging}) and (\ref{ec:survival}). These equations
use background values of concentrations and environmental degradation,
that are continuous fields defined on the biofilm, governed by the diffusion
problems (\ref{ec:oxygen}), (\ref{ec:antibioticbio}) and (\ref{ec:degradation}).

The computational strategy is the following:
\begin{itemize}
\item For a fixed biofilm configuration,
we let the diffusion problems for $C_o$, $C_a$ and $\varepsilon$ relax to
stationary values, which occurs in a short time scale $\tau$ (seconds). These variables  are  then defined  and fixed in all the control volumes. 
\item For each cell in the biofilm, we update the values of the variables
$e$, $v$, $v_e$, $[C_{IN}]$, $a$, $h$, $q$, $p$  solving equations (\ref{ec:scaledenergy1}), (\ref{ec:scaledcellvolume1}), 
(\ref{ec:scaledepsvolume1}), (\ref{ec:antibiotic}), (\ref{ec:acclimation}),
(\ref{ec:hazard}), (\ref{ec:aging}) and (\ref{ec:survival}) in a slower time
scale $T$ (minutes) using the background values of 
$C_o$, $C_a$ and $\varepsilon$.
\item We revise the status of all cells in the biofilm, in a random ordering.
Cells differentiate and divide according to the rules stated in Section
\ref{sec:division}. To decide whether a cell dies, we generate a random
number $n\in (0,1)$ and kill the cell if $p < n {N_a \over N},$ where $N$ 
is the total number of cells and $N_a$ the current number of alive cells. 
\item We update the concentration fields and repeat the procedure.
\end{itemize}

\begin{figure}[h!]
\centering
\includegraphics[width=12cm]{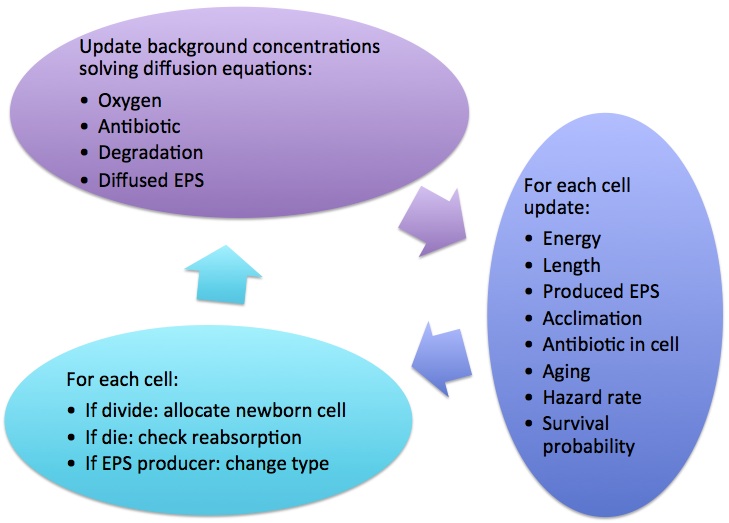}
\caption{{\bf Schematic description of the computational arrangement.}
Continuous equations for concentrations are discretized in the square grid
representing the biofilm.
Each tile contains cells of certain volume and energy,
with an associated EPS volume around it, whose dynamics is governed 
by the DEB equations. Here, we assume that each tile contains just one 
cell. In practice, all these items are stuck together in a more
amorphous mass. This is a computational idealization allowing us to combine
macroscopic and microscopic information.} 
\label{fig6}
\end{figure}

Allocating newborn cells, reabsorbing dead cells and allocating produced
EPS in a 2D or 3D biofilm geometry are challenging issues for which many
ideas have been proposed depending on the biofilm types \cite{prebacillus,scirep,allen,ibmfluid,kapellos,ibmsolid}. 
We do not intend to address these issues
here. Instead, we work in a simplified geometry, implementing simple rules
that allow us to check the performance of the DEB dynamics in a spatially
varying film. In the numerical simulations presented here we make
the following assumptions:

\begin{itemize}
\item The biofilm is a 2D region, divided in cubic tiles for computational
purposes.
\item Each tile contains one cell together with EPS and dissolved substances.
The idea is schematically represented in Figure \ref{fig5}. 
\item When a cell divides, the length and energy split in two. The daughter
cell shifts other cells in the direction of minimal mechanical resistance (closest
to the surface or to a dead cell).
\item When a cell dies, it is reabsorbed by its neighbors, provided there are
enough neighbors alive. A neighbor or newborn cell occupies its place in that 
case. Otherwise, its stays there and we get a necrotic region.
\item A fraction $\alpha \in (0,1)$ of the EPS produced by equation (\ref{ec:scaledepsvolume1})
remains attached to the cell. The rest diffuses through the whole biofilm.
Therefore, when updating the concentrations $C_a$, $C_o$, $\varepsilon$,
we need to update the diffused concentration of EPS
$C_e(\mathbf x, t)$ too:
\begin{eqnarray}
{\partial C_e\over \partial t}
= d_e \Delta C_e +  (1-\alpha) R_e X,
\label{ec:difeps}
\end{eqnarray}
with zero flux boundary conditions ${\partial C_e \over \partial {\bf n}}=0$
on the biofilm boundary.  
The equations in Sections \ref{sec:oxygen}-\ref{sec:effect}
set $\alpha=1$. All the EPS produced by a cell remains attached to it.
When $\alpha \neq 1$, $R_e$ is replaced by $\alpha R_e + R_e'$
in all these equations.
$\alpha R_e$ represents the fraction of EPS produced at each site that
remains in it. It may be zero when there are no EPS producers in that
tile. $R_e'$ represents the fraction of EPS accumulated due to global
diffusion processes governed by (\ref{ec:difeps}). It will always be
non zero by diffusion. The way we evaluate $R_e'$ is the following.
If at a certain control volume we start with an initial concentration 
$C_e(t_1)$ to reach a final concentration $C_e(t_2)$ this defines a rate $R_e'={1\over t_2-t_1}\ln ({C_e(t_2) \over C_e(t_1)}).$ 
The overall computational chart is represented in Figure \ref{fig6}.
\end{itemize}

The physical justification for this treatment of EPS is that the EPS matrix 
is formed by polymers. 
Monomers will diffuse easily, but a fraction of them will form polymeric 
chains of increasing size attached to the cells.
An empirical justification is that if the produced EPS remains attached to producers, the active cells not producing EPS in outer layers die too fast
and form unphysical necrotic layers in absence of any toxicants. In practice, the EPS matrix envelops and shelters all the biofilm cells.

\section{Parameter calibration}
\label{sec:parameter}

\begin{table} 
\hskip -1.5cm
\begin{tabular}{|c|c|c|c|c|}
\hline 
Symbol & Units & Name & Value & Source \\
\hline 
$\rho_x$  & ${{\rm mg} \over \ell} $& cell density & $47000$  & \cite{stewartmodel} \\
 & $\mu$m & biofilm size & $1000 \times 200$ & chosen \\
$L$ & $\mu$m & cell length & $1.67-2.75$ & \cite{robinsonepsrates} \\
$2\rho$ & $\mu$m & cell diameter & $0.5-0.73$ & \cite{robinsonepsrates} \\
$C_{o,out} $  &  ${{\rm mg} \over \ell} $ & Oxygen concentration & $0.035$ & \cite{stewartmodel} \\
$K_o$ & ${{\rm mg} \over \ell}$ & Oxygen half saturation &  $0.1$  & \cite{stewartmodel} \\
$d_o $  &  ${\mu {\rm m}^2 \over {\rm s}} $ & Oxygen diffusion & $2.2 \times
10^{4}$  &
\cite{werneroxygen}\\
$\mu_{max}$ & ${\rm hr}^{-1}$ & Growth rate with oxygen & $0.3$ &  
\cite{stewartmodel} \\
$Y_{x/o}$ & ${{\rm mg \, cell} \over {\rm mg \, oxygen}}$  & cell yield & $0.24$ 
&  \cite{stewartmodel} \\
$k$ & ${{\rm mg \, polymer} \over {\rm mg \, cell}}$
&  growth associated yield & $2.2371$ &  estimated \\
$k'$ & ${{\rm mg \, polymer}  \over {\rm mg \, cell \, hour}}$
&  non growth associated yield & $0.29$ &  estimated   \\
$Y_{x/o}$ & ${{\rm mg \, cell}  \over {\rm mg \, oxygen} }$
&  cell yield & $ 0.34 $ & estimated\\
$Y_{p/o}$ & ${{\rm mg \, polymer}  \over {\rm mg \, oxygen}  }$
&  polymer yield & $  0.56 $ &  estimated  \\
$C_{a,out} $  &  ${{\rm mg} \over \ell} $ & Antibiotic concentration & 
$0.78, 0.20, 3.13,1.56$
 & \cite{ishidadata} \\
$d_a$  &  ${\mu {\rm m}^2 \over {\rm s}} $ & Antibiotic diffusion & 
$0.5 \times 10^4$ & \cite{stewartmodel} \\
$d_e$  &  ${\mu {\rm m}^2 \over {\rm s}} $ & EPS diffusion & $10^4$ 
& estimated  \\
$d_\varepsilon$  &  ${\mu {\rm m}^2 \over {\rm s}} $ & Degradation diffusion 
&  $0.5 \times 10^4$ &   estimated \\
$\nu$  & ${\rm hr}^{-1}$   & Energy conductance & 0.84768  & \cite{klanjscekplos}\\
$m_\kappa$  & ${\rm hr}^{-1}$     &  Maintenance rate & 0.1266  &  estimated  \\
$g_\kappa $    & n.d.      &  Investment ratio   & 0.9766 & estimated \\
$s_q$ & ${\ell \over {\rm mg}}$ & Multiplicative stress coeff.  & $0.8921 \times 10^{-5}$  &  estimated \\
$h_a$  & ${\rm hr}^{-2}$    & Weibull aging acceleration & $ 1.4192
\times  10^{-4}  $  & estimated \\
$h_0 $  & ${\rm hr}^{-1}$    & Initial hazard rate & 0.4  &  estimated  \\
$k_{qA}^I$ & ${{\rm Cmol} \over {\rm mg \, hour}^3} $ & Dissolved antibiotic toxicity  &  $ {s \over C_{a,out}}  {k_A^O \over k_A^I}$ & estimated \\
$k_A^I$ &  ${ \ell \over {\rm hr \, Cmol}}$ & Antibiotic influx coeff. &  
$8.6 \times 10^{-6} $ & \cite{klanjscekplos} \\
$k_A^O$ & ${\rm hr}^{-1}$  & Antibiotic efflux coeff.  &  0.17251 &  \cite{klanjscekplos}\\
$K_V$ & ${{\rm mg} \over \ell}$  & Noncompetitive inhibition coeff. & 154.82   & \cite{klanjscekplos} \\
$a_M$ & n.d. & Target acclimation energy & 1.6703 &  \cite{klanjscekplos}\\
$\nu_{\varepsilon} $& ${\ell \over {\rm mg}}$ & Environmental degradation coeff. &
$0.23566/12000 $ &  \cite{klanjscekplos}\\
$\nu_m$ & n.d. & Maintenance respiratory coefficient & $0.054703$ &  \cite{klanjscekplos}\\
$\gamma_\varepsilon $ & n.d. & Environmental degradation effect coeff. & $1$ &  \cite{klanjscekplos} \\
\hline
\end{tabular}
\caption{{\bf Parameter values used in the simulations.} Recall that $1$Cmol
 = $12$g.}
\label{table2}
\end{table}

Table \ref{table2} lists the parameters used in the simulations.  
The diffusion coefficient for oxygen $d_o$
is selected so that oxygen penetration inside the biofilm
ranges in the values reported  
(about $50$ $\mu m$) \cite{werneroxygen}. 
Oxygen penetration depth is defined
as the distance into the biofilm at which the first derivative
of the oxygen concentration reaches $5\%$ of its maximum
value. As said before, oxygen is considered to be the limiting factor \cite{hoibyresistance} for biofilm development in medical environments. 
We lack measurements of $k$, $k'$, $Y_{x/o}$ and $Y_{p/o}$
in this case, but we may propose values for 
$Y_{x/o}$ and $Y_{p/o}$, $k$ and $k'$ based on an educated
guess respecting the proportions 
observed for carbon \cite{robinsonepsrates}. 

Values for the maintenance rates $m_\kappa$ and investment 
ratios $g_\kappa$ are inferred as follows. 
The macroscopic equations for EPS and cell production are
\cite{robinsonepsrates}:
\[
{dX_e \over dt}= [ k \mu_{max} f + k' ] X,
\quad {dX \over dt}= \mu_{max} f X.
\]
In absence of acclimation, toxicity and harzard corrections,
that is $a=a_M$, $\nu'=\nu$, $h=0$,
equations (\ref{ec:scaledcellvolume1}) and  (\ref{ec:scaledepsvolume1}) 
read
\[
{dv_e \over dt} =  \left( {g_{\kappa}\over g_{e,\kappa}} r 
+ {m_{\kappa}\over g_{e,\kappa}}\right) v, \quad {dv \over dt} = rv, \quad
r= \left({\nu e - m_{\kappa}  g_{\kappa}
\over e+g_{\kappa} }\right)^+.
\]
Notice that the energy governed by (\ref{ec:scaledenergy1}) tends
to an equilibrium $e=f$ for each cell. We set $f=f_{max} ={ C_{o,out} \over C_{o,out} + K_o} $ and $e=f_{max}$. Dimensions can be restored
in these equations by scaling $v$ and $v_e$.
Comparing with the equations for $X_e$ and $X$, we approximate
${g_\kappa \over m_\kappa}$ by ${k\over k'}$ and
${\nu f_{max} - m_\kappa g_\kappa \over
f_{max} + g_\kappa}$ by $\mu_{\max}  f_{max}$.
Taking $m_\kappa = {k'\over k} g_\kappa$, $g_{\kappa}$
is a positive solution of a second order equation, given
by
\[
g_\kappa = {1\over 2} \left( -{k\over k'} \mu_{max} f_{max}+ \sqrt{
\big( {k\over k'} \mu_{max} f_{max} \big)^2 - 4 {k \over k'}
(\mu_{max}  f_{max} - \nu) f_{max}} \right).
\] 

The values of the multiplicative stress coefficient $s_q$, Weibull aging 
acceleration $h_a$, dissolved antibiotic toxicity $k_{q,A}^I$, influx and 
efflux coefficients $k_A^I$, $k_A^O$, and the initial hazard rate $h_0$
in equations (\ref{ec:hazard}), (\ref{ec:aging}), (\ref{ec:agingacclimation}) and (\ref{ec:survival})
are fitted to data in reference \cite{ishidadata}. To do so, we
consider a single cell of volume $v$, energy $e$ with 
density $\rho_x$ and write down the equations for the
hazard rate $h$, aging acceleration $q$ and survival probability
$p$ setting $[C_{IN}]=C_{a,out} {k_A^I \over k_A^O},$ the equilibrium
value. Taking $a=h_{0}$, $b=r+r_e$, $c=e\,h_a (\nu'-r)$,
$d=e\,s_q\,\rho_x v (\nu'-r)$ and $s=k_{qA}^I [C_{IN}]$, the  solutions 
are:
\[
q(t)={c+s \over d-b}\left(e^{(d-b)t}-1\right),
\]
\[
h(t)=\left(a-{c+s \over d-b}\left({1 \over d}-{1 \over b}
\right)\right)e^{-b\,t}+{c+s \over d-b}\left({e^{(d-b)t} \over
d}-{1 \over b} \right),
\]
\begin{eqnarray}
p(t)=exp\Bigg({a \over b}\left( e^{-b\,t}-1 \right)-{c+s \over
d-b}\Bigg(\Big({1 \over b\,d}-{1 \over b^2}
\Big)e^{-b\,t}+ \nonumber \\
{e^{(d-b)t} \over d (d-b)}-{1 \over
b}t-{1 \over b\,d}+{1 \over b^2} -{1 \over
d(d-b)}\Bigg)\Bigg).
\label{pexplicit}
\end{eqnarray}

\begin{figure}[h!]
\centering
\includegraphics[width=6.5cm]{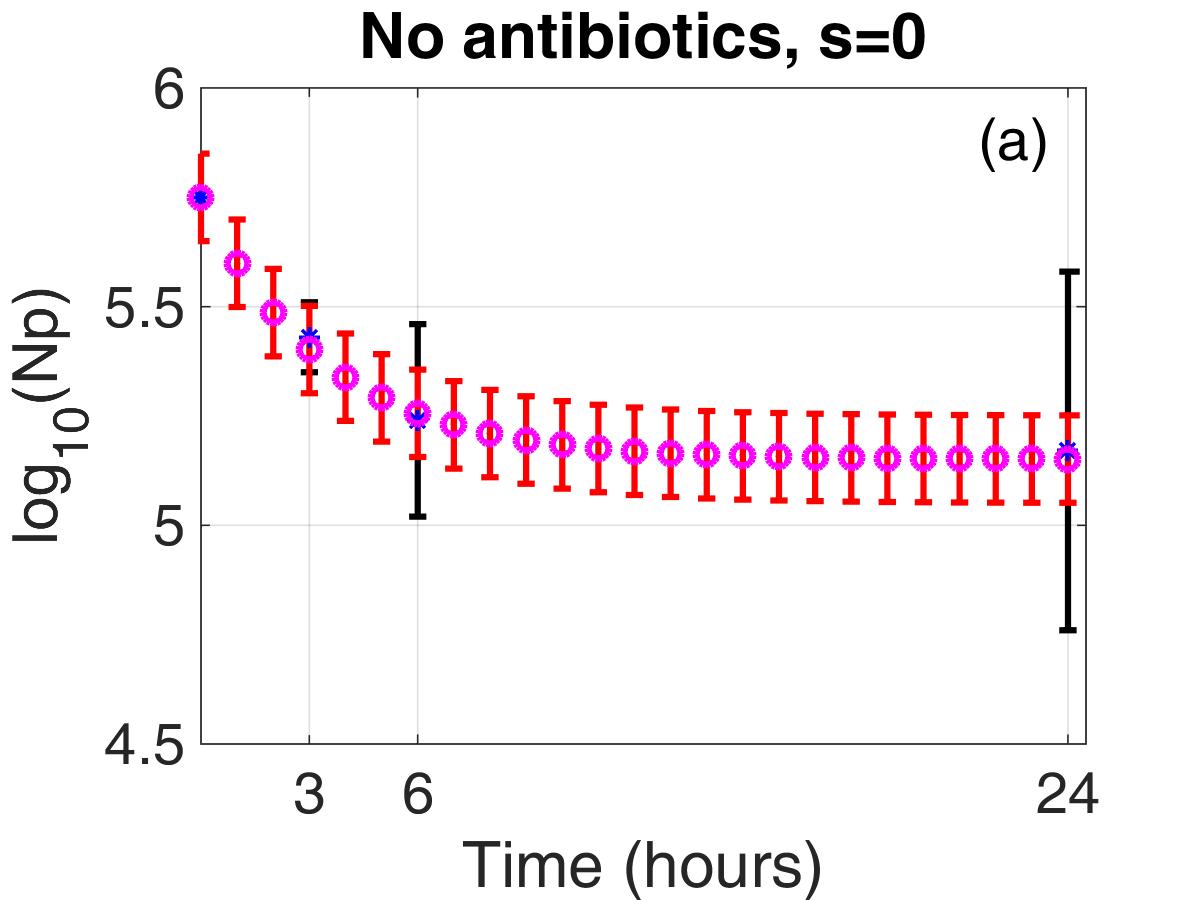}\\
\includegraphics[width=6.5cm]{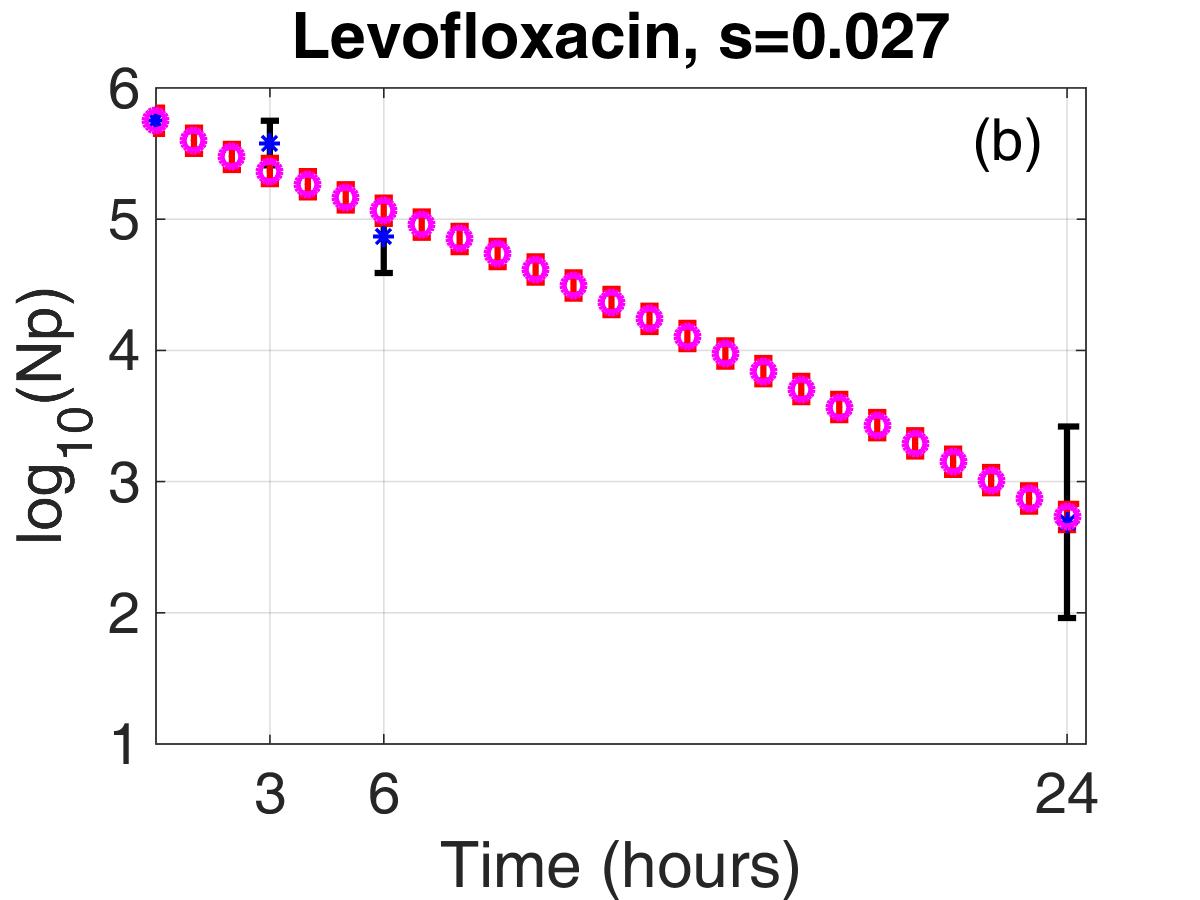}
\includegraphics[width=6.5cm]{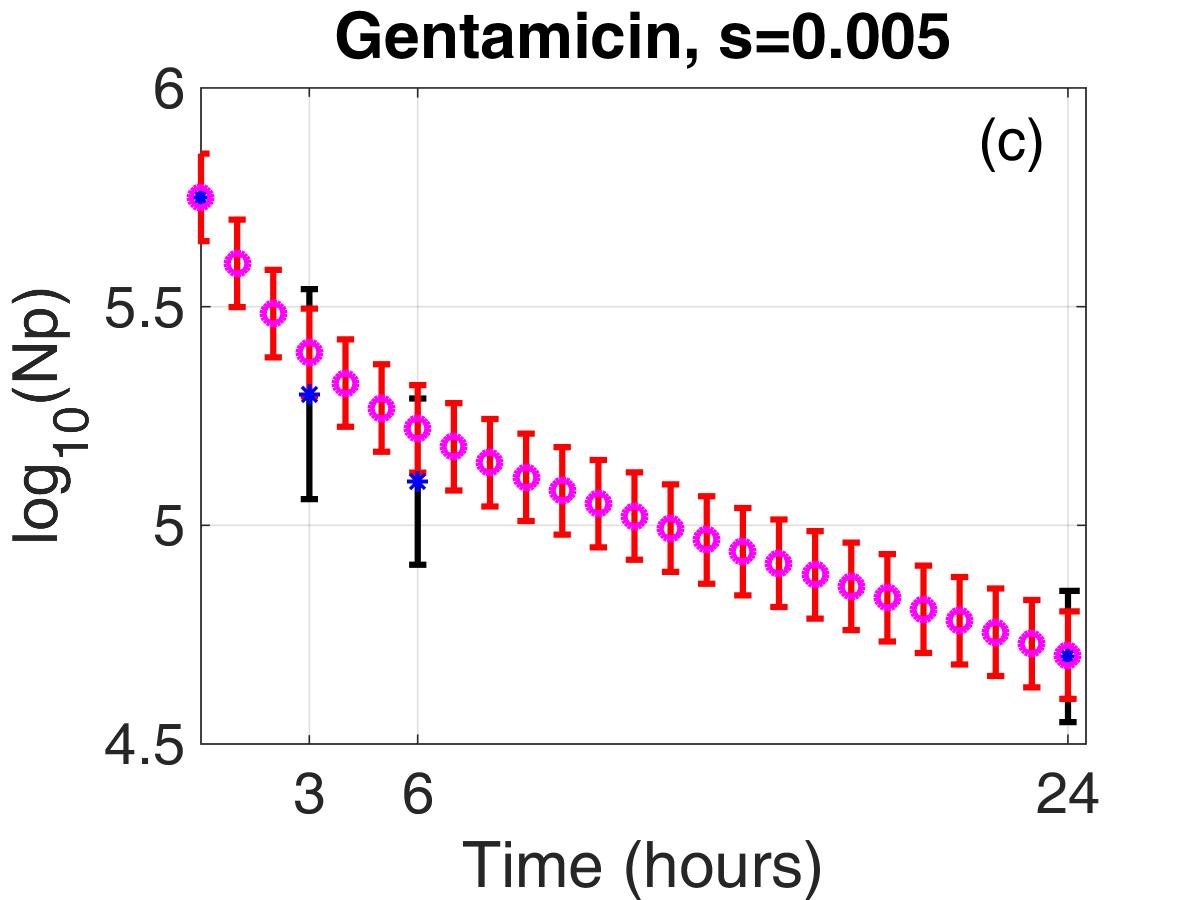} \\
\includegraphics[width=6.5cm]{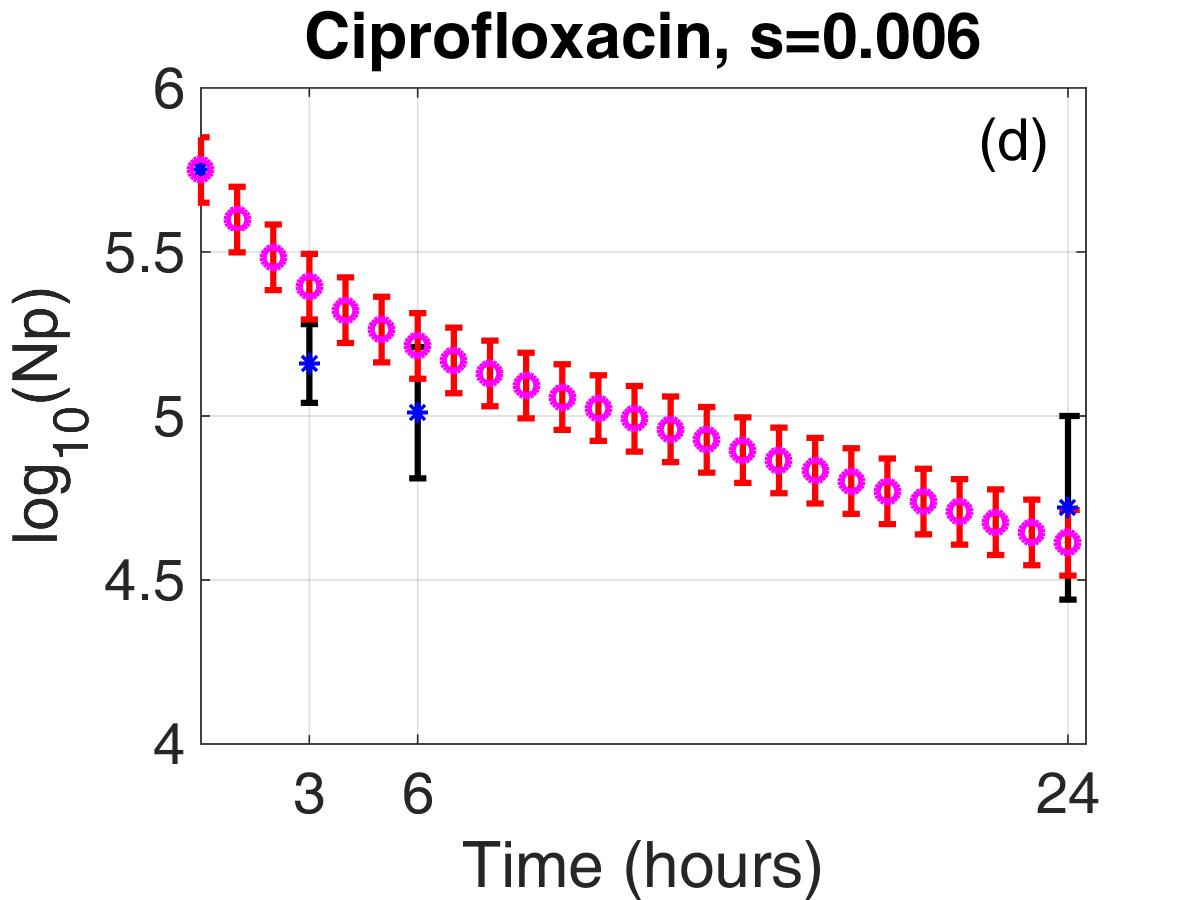}
\includegraphics[width=6.5cm]{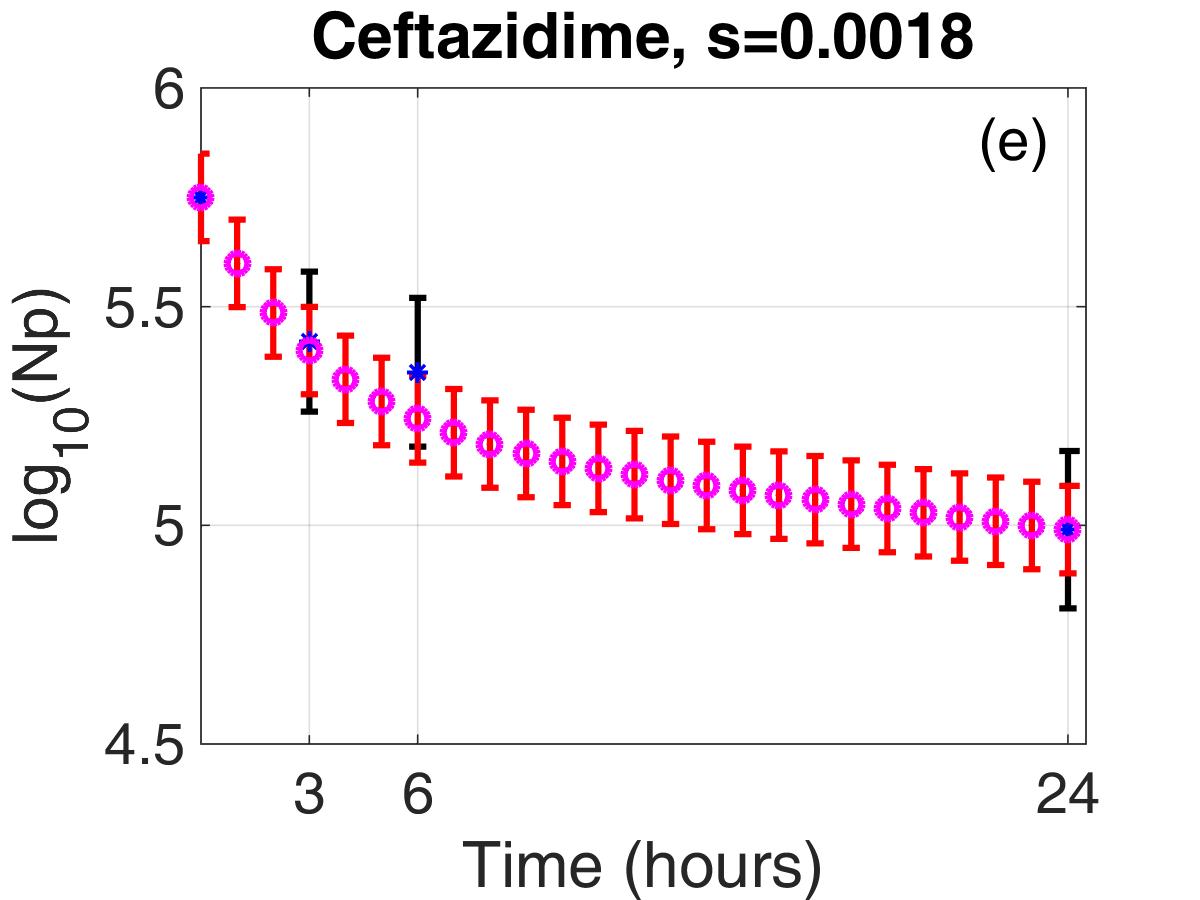}
\caption{{\bf Comparison of theoretical predictions and experimental
data.} The predictions provided by formula (\ref{pexplicit}) for $p$ are 
compared to the data in Table 1 of Reference \cite{ishidadata}.
The parameter values for $p(t)$ are $a=0.4$, $b=0.29$, 
 $c=3.61 \times 10^{-7}$ and $d=4.8 \times 10^{-3}$. 
The additional parameter $s$ is adjusted for each antibiotic.
Circles represent $log_{10}(p\cdot N)$ with $N=10^{5.75}$ 
(magenta) and $N=10^{5.75\pm 0.10}$ (red). 
Asterisks represent the measurements in Table 1 
of  Reference \cite{ishidadata} for the same antibiotic
doses. As before, light (blue) colors correspond to the mean
measured value, whereas dark (black) colors
mark the error interval.}
\label{fig7}
\end{figure}

The way we find values for $a$, $b$, $c$, $d$, $s$ is the following. We first remove the antibiotic and seek to adjust the viable cell counts in Table 1 of reference \cite{ishidadata} for the control population. Figure \ref{fig7}
compares the bacterial counts at times $3$, $6$, $24$ hours with our fittings of $log_{10}(p N)$ for a population of the initial size $N=10^{5.75}$ CFU/m.
Once these parameters are fitted, we successfully fit the remaining
parameters to the approximate 
number of viable cells for four different antibiotics: levofloxacin, ciprofloxacin, gentamicin, ceftazidime.

We approximate the data in Table 1 of  \cite{ishidadata} in absence of 
antibiotics dividing by $10$ the values  $s_q$ and $h_a$ from reference \cite{klanjscekplos}. We choose average values for $e$ and $v$,
 $v=0.5$,  $e = median(f(C_{o,out}))\approx 0.003$, and set
$\nu'=\nu$. Then,  $c=e\,h_a (\nu'-r) \approx 3.61 \times 10^{-7}$ 
y $d=e\,s_q\,\rho_x v (\nu'-r) \approx 4.8 \times 10^{-3}$. 
We vary $a$ and $b$ so as to approximate the data in Table 1 
of \cite{ishidadata}: $a=h_0=0.4$ and $b=0.29$. 
These values are chosen because the fit  still holds when
we replace single cells by a biofilm formed by a few hundred
layers. Since $r\sim 0$ for EPS cells, we set $k'=b$ and keep
the ratio ${k\over k'}$ observed for carbon.

\section{Numerical results}
\label{sec:results}

\begin{figure}
\centering
\includegraphics[width=10cm]{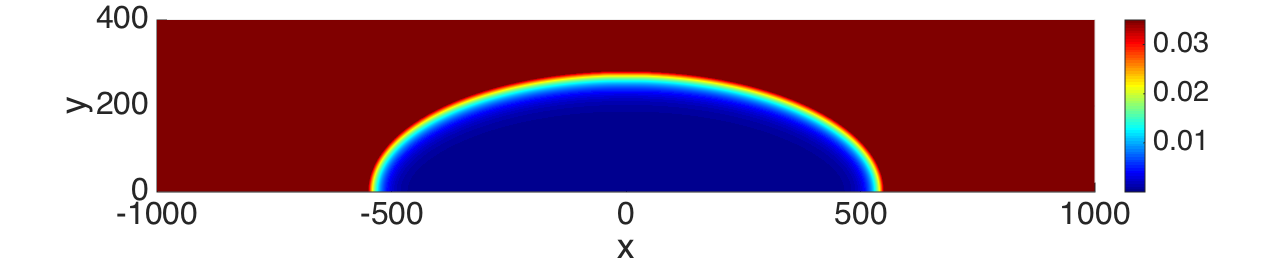}
\caption{\bf Equilibrium oxygen concentration in dimensionless
units ${C_o \over K_o}$.} 
\label{fig8}
\end{figure}

The purpose of this section is to investigate the influence of the spatial 
variations of the concentrations in cell death.
To simplify, in the small scale 2D simulations  presented here we 
consider tiles of size $1 \, \mu m \times 1 \, \mu m$ in
Figure \ref{fig5}. 
Figure \ref{fig8} shows the spatial structure of the oxygen distribution in a
biofilm  of maximum width $1000 \, \mu m$ and maximum height $200 \, \mu m.$
We solve equation (\ref{ec:oxygen}) discretized in the grid, with Dirichlet boundary condition $C_o=C_{o,out}$ on the interface with air and Neumann boundary condition ${d C_o \over d n}=0$ on the bottom interface. We take 
 $f(C_{o})= {C_{o} \over C_{o} +K_o}$
and $X= \rho_x v$, where $v$ is the dimensionless volume of the
cell contained in the tile. Figure \ref{fig8} sets $v=0.5$.
Initially, $C_o=0$ inside the biofilm. We solve using a explicit
scheme until $C_o$ relaxes to a stationary configuration:  
\[
C^{\ell+1}_{o;i,j}=C^{\ell}_{o;i,j} + \delta t \, d_o 
{ C^{\ell}_{o;i,j+1} + C^{\ell}_{o;i+1,j} -4 C^{\ell}_{o;i,j} + C^{\ell}_{o;i,j-1} 
+ C^{\ell}_{o;i-1,j}  \over \delta x^2 } + \delta t \, g(X_{i,j}),
\]
where $g(X)$ represents the source in equation (\ref{ec:oxygen}).
The spatial step $\delta x$ is the side of a tile. Tiles in the spatial grid
are labelled using a couple of indices $(i,j)$ to indicate their position.
Time is discretized with time step $\delta t$, so that $t_{\ell+1}=
t_{\ell} +\delta t =(\ell +1)  \delta t$ and $C_{o;i,j}^\ell \sim 
C_{o;i,j}(t_\ell)$. For this scheme to be stable,
we need ${\delta t \, d_o \over \delta x^2} \leq  0.5$. We advance
in a time scale of seconds. Similar schemes are used to update the
additional space dependent concentrations $C_a$, $C_{\varepsilon}$,
$C_e$ when needed. 

\begin{figure}
\centering
\includegraphics[width=10cm]{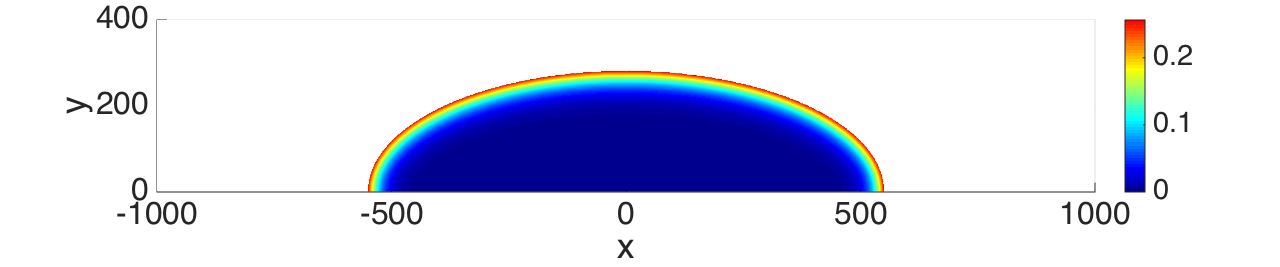}
\caption{\bf Equilibrium energy distribution} 
\label{fig9}
\end{figure}

The cell energies are governed by DEB equations depending on 
$f={C_o \over C_o + K_o}$, which varies through the biofilm. 
Figure \ref{fig9}  represents the 
equilibrium distribution $e=f(C_o)$ at each tile. 
The maximum energy is $e_m={C_{o,out} \over C_{o,out} +K_o}.$
In view of the energy distribution, we set threshold values
$e_{eps1} = 7 \times 10^{-6}$, $e_{eps2} = 0.02$ and $e_d=0.01$
 to produce EPS or divide. 

\begin{figure}
\centering
(a) \\
\includegraphics[width=10cm]{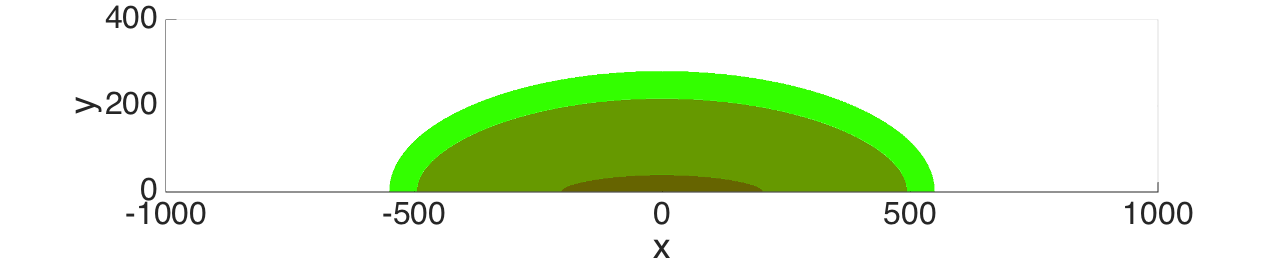} \\
(b) \\
\includegraphics[width=10cm]{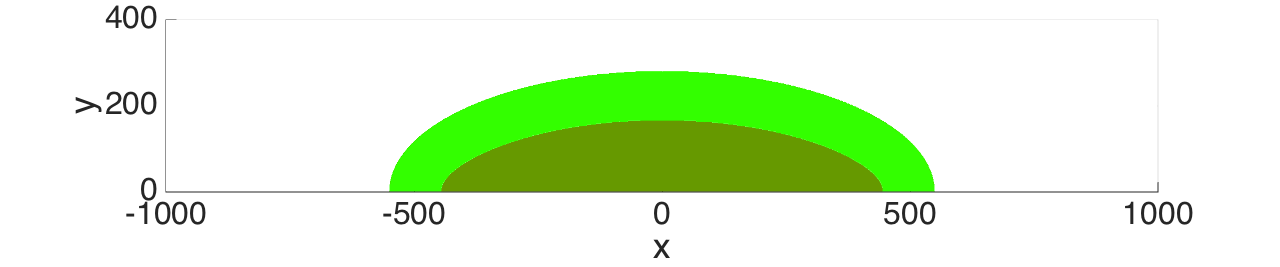}
\caption{{\bf Equilibrium distribution of cell types}: (a) for the equilibrium
energy $e=f(C_o)$, (b) for larger energy $e=10 f(C_o)$. Light green
represents active cells. Dark green represents dormant cells.
Green represents EPS producers.} 
\label{fig10}
\end{figure}

Starting from a uniform distribution of cell types (non EPS producers, $v_e=0$)
with volumes $v(0)=0.5$  we compute the evolution of their energy and size to check the evolution of the cell type distribution with time. We choose as initial energy a  perturbation of the equilibrium energy distribution at each location, 
and fix  $q>q_{eps}=10^{-8}.$
By solving the equations for $e$, $v$, $v_e$ for each cell tile ignoring
aging and hazard, we see that  cells evolve to be EPS producers in an intermediate region. The outer layers are normal active cells, likely to divide. 
A thin inner core may be formed by cells with little energy than do not divide and 
do not produce EPS. The equilibrium distribution of types is depicted in Figure \ref{fig10}(a). If we increase the initial energy, we may find intermediate
states such those in Figure \ref{fig10}(b), where that thin inner layer vanishes.
Repeating the simulation with initial volumes $v$ randomly distributed between $(0.45,0.6)$  and random perturbations of the equilibrium energy we find
similar results with a smoother transition between regions, reflecting the noise.

\begin{figure}
\centering
\includegraphics[width=6.5cm]{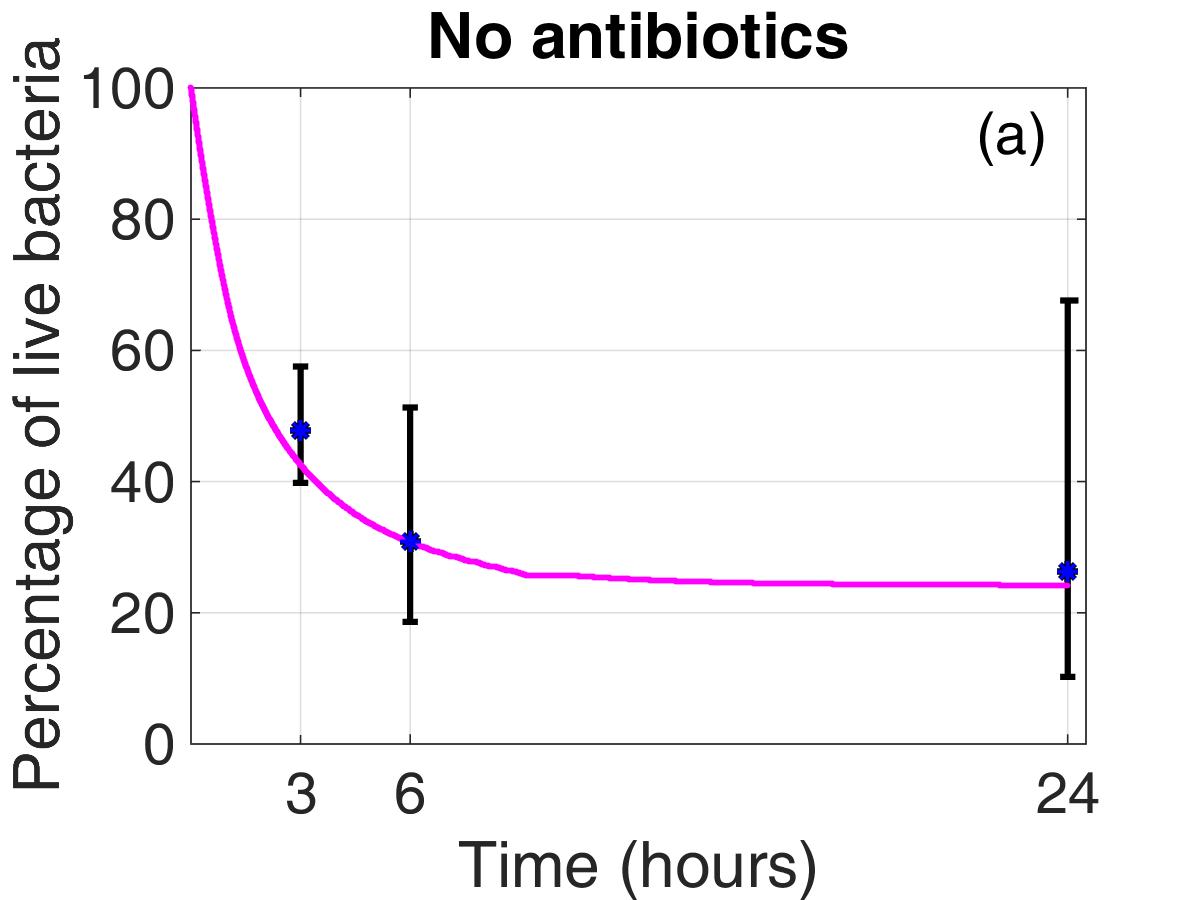}
\includegraphics[width=6.5cm]{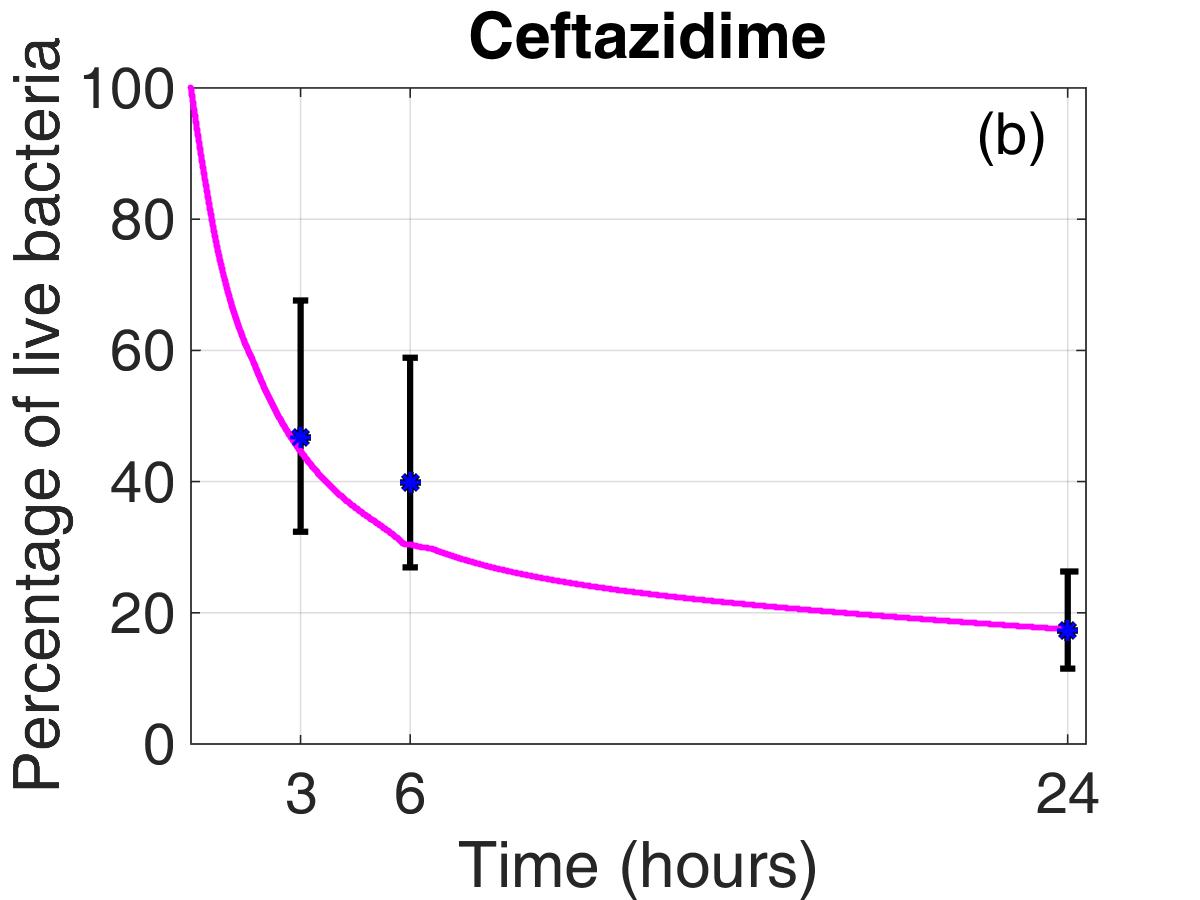}
\caption{{\bf Simulated live cell counts compared to experimental
measurements} in Table
1 of Reference  \cite{ishidadata}: (a) in absence of antibiotics, (b) for 
ceftazidime.  The continuous curve represents the simulated values.
Asterisks depict the measurements in Table 1 
of  reference \cite{ishidadata} for the same antibiotic
doses, black lines represent the error bars.
} 
\label{fig11}
\end{figure}

To study the distribution and number of dead cells, as well as the effect of 
antibiotics, we activate the equations for aging and hazard rates. We 
compute a reference concentration of antibiotic $C_a$ in the biofilm solving (\ref{ec:antibioticbio}) with Dirichlet boundary condition $C_a=C_{a,out}$ on  the interface biofilm/air and Neumann boundary condition
${d C_a \over d n}=0$ on the bottom interface. We assume that all cells are
undifferentiated in this reference computation ($R_e=0$) and employ
the equilibrium oxygen and energy distributions to evaluate $R$.
The concentration of antibiotics inside the biofilm evolved towards a 
constant, and we will take it to be approximately constant and
equal to $C_{a,out}$ in the biofilm. Figure \ref{fig11} shows that the fittings 
of the time evolution of the survival rates performed in Section \ref{sec:computational} persist when all the biofilm is taken into account.
A low energy cell population is considered in that figure, forbidding cell
division for simplification.

\begin{figure}
\centering
(a) \\
\includegraphics[width=10cm]{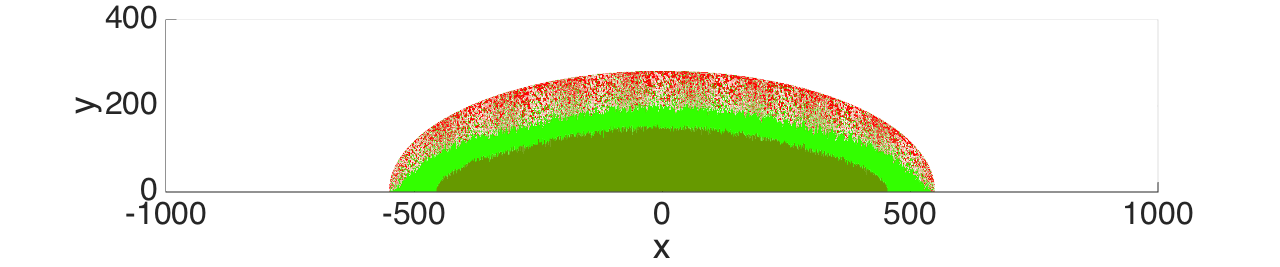} \\
(b) \\
\includegraphics[width=10cm]{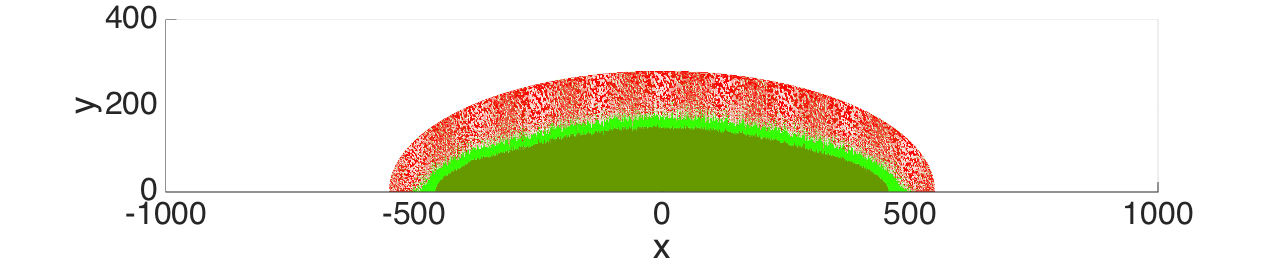} \\
(c) \\
\includegraphics[width=10cm]{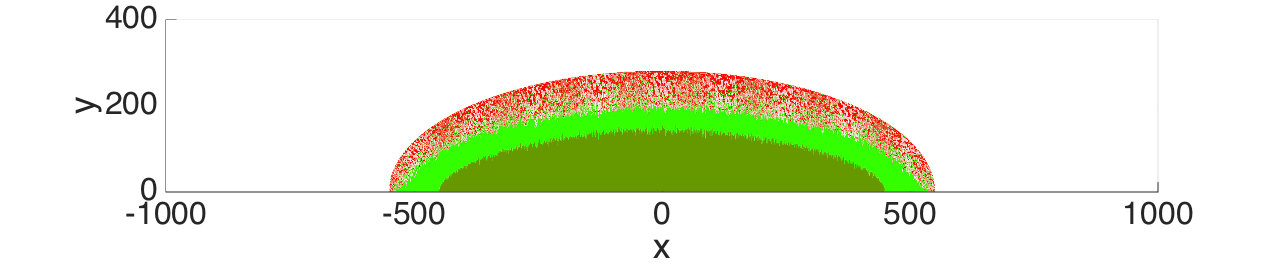}
\caption{{\bf Distribution of cell types applying Ceftazidime}.
(a) for the reference dose $C_a=3.13 \, {\mu g \over m\ell}$ 
after 90 minutes,  (b) same dose, after 180 minutes, 
(c) a dose $100$ fold larger, after 90 minutes. 
Dead and active cells are depicted in red and green colors,
whereas EPS producers are dark green.
The number of dead cells increase with time and the doses.
The initial energy is $e(0)=10 f(C_o)$.} 
\label{fig12}
\end{figure}

The model yields information on qualitative behavior that agrees with
some experimentally
observed tendencies and may be useful to understand the
causes or to predict new behaviors. Panels (a) and (c) in Figure
\ref{fig12} reveal that for an antibiotic acting mostly on active cells
like Ceftazidime (it inhibits cell wall synthesis), cells mostly die in the
outer biofilm layer and   the extent of the dead region increases
with the dose of  antibiotic. Panel (b) shows that it also increases
with time.  We are assuming that the antibiotic has diffused inside
the biofilm so that the spatial distribution of dead regions depends
on what type of cells are preferential targets of the antibiotic under
study. For antibiotics like Ceftazidime, acting on active cells,
we use toxicity coefficients depending on the cell activity, 
 of the form  ${e \over 0.1 \, {\rm median}(e)}k_{qA}^I$. 
For other  antibiotic like colistin, targeting
the dormant core cells, we may use toxicity coefficients
decreasing with the cell activity instead.
Mutations increasing the efflux might be accounted for
raising the value of that coefficient. Augmented resistance
to deactivating enzymes may be incorporated modifying the
toxicity.

\begin{figure}
\centering
(a) \\
\includegraphics[width=10cm]{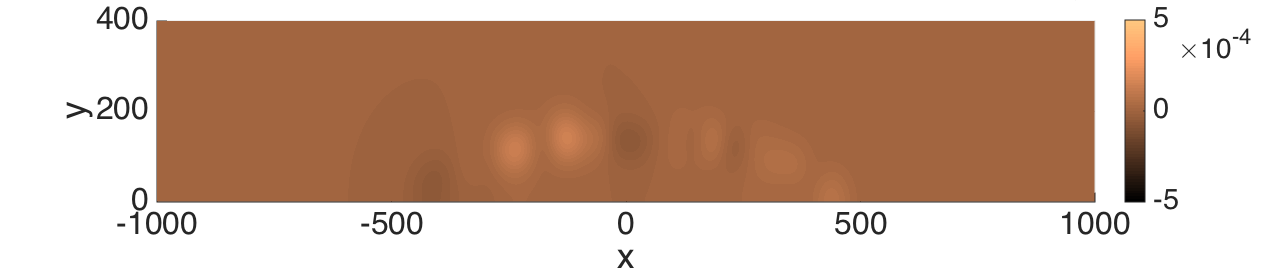} \\
(b) \\
\includegraphics[width=10cm]{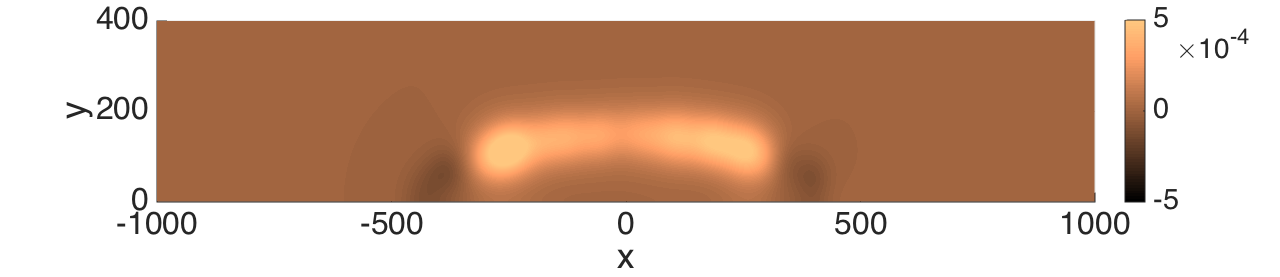} \\
\caption{{\bf Difference $C_e-C_e^{(1)}$ of EPS concentrations when 
antibiotics are administered}. $C_e^{(1)}$ represents the concentration
when $C_{a,out}=3.13 \, {\mu g \over m\ell}$.
(a) $C_e-C_e^{(1)}$, $C_e$ computed for 
$C_{a,out}=6.26 \, {\mu g \over m\ell}$. 
(b) $C_e-C_e^{(1)}$, $C_e$ computed for 
$C_{a,out}=18.78 \, {\mu g \over m\ell}$. 
The initial energy is $e(0)=10 f(C_o)$ and the difference is
represented after one hour in both cases.
The mean value of the differences is positive.}
\label{fig14}
\end{figure}

In absence of antibiotics cells die in the inner
regions in a scattered way.  Active cells in the outer layers
are still more likely to die, a standard fact in the DEB theory, that
assumes organisms with more energy more likely to suffer
damage. That effect would be diminished increasing the
difussivity of produced EPS $d_e$, or allowing differentiation
into EPS producers for larger energies.
Cells that die in the outer layers in absence of antibiotics
may also be eroded by external agents or absorbed by newly 
born cells. Unlike the previous case, we do not   get an
expanding outer necrotic region. Analyzing the amount of
EPS produced with and without antibiotics, we find that 
the presence of antibiotics enhances EPS production,
see Figure \ref{fig14}. This fact is also responsible for the
small difference between panels (a) and (c) of Figure \ref{fig12}.
The number of dead cells is larger for panel (c), but not
as large as could be expected from the dose increase due
to enhanced EPS production that neutralizes that fact.

\section{Conclusions}
\label{sec:conclusions}

Coupling dynamic energy budget dynamics for the metabolism of individual cells to diffusion equations for the evolution of chemical fields we have reproduced qualitative trends experimentally observed in bacterial biofilms under the action of antibiotics. We calibrate the parameters for a quantitative agreement with the death rates measured for {\em P. aureginosa} under a number of antibiotics and
study spatial variations in the distribution of dead and alive cells. Our numerical simulations show that the presence of antibiotics enhances  production of polymeric matrix, as expected from true measurements. For antibiotics targeting active cells, we observe the formation of a necrotic outer layer, that expands deeper in the biofilm as time goes on and the antibiotic dose increases. Our tests suggest ways to handle antibiotics targeting dormant cells and resistance mechanisms such as efflux pumps or enzymes. New experimental measurements \cite{naganoinfluxeffluxec} would be required to calibrate the mechanisms to practical cases. 
The current computational model enables validation and hypothesis testing when developing therapies to handle chronic infections caused by biofilms.
Determining the true relevance of increasing the dose and whether periodic or continuous infusion of antibiotics is more effective \cite {kimdoses}  would be essential for practical applications.

\vskip 5mm
{\bf Acknowledgements.} Research partially supported by the NILS Mobility Program  and MINECO grant No. MTM2014-56948-C2-1-P.\\



\begin{thebibliography}{9}

\bibitem{abbasmbic} H.A. Abbas, F.M. Serry, E.M. EL-Masry, Combating {\em Pseudomonas aeruginosa} biofilms by potential biofilm inhibitors, Asian J. Res. Pharm. Sci. 2, 66-72, 2012.
\bibitem{alipourdissolve} M. Alipour, Z.E. Suntres, A. Omri, Importance of DNase and alginate lyase for enhancing free and liposome encapsulated aminoglycoside activity against Pseudomonas aeruginosa. J Antimicrob Chemother 64, 317-325, 2009.
\bibitem{anwarenhanced} H. Anwar, J.W. Costerton, Enhanced activity of combination of tobramycin and piperacillin for eradication of sessile biofilm cells of {\em Pseudomonas aeruginosa}, Antimicrob Agents Chemother 34, 1666-1671, 1990.
\bibitem{baggeincreasedeps} N. Bagge, M. Schuster, M. Hentzer, O. Ciofu, M. Givskov, E.P. Greenberg, et al, {\em Pseudomonas aeruginosa} biofilms exposed to imipenem exhibit changes in global gene expression and $\beta$-lactamase and alginate production.
Antimicrob. Agents Chemother. 48, 1175-1187, 2004.
\bibitem{debeeroxygen} D. de Beer, P. Stoodley, F. Roe, Z. Lewandowski, Effects of biofilm structure on oxygen distribution and mass transport, Biotechnol Bioeng 43, 1131-1138, 1994.
\bibitem{broounresistance} A. Brooun, S. Liu, K. Lewis, A dose-response study of antibiotic resistance in {\em Pseudomonas Aeruginosa} biofilms, Antimicrob. Agents Chemother.  44, 640-646, 2000.
\bibitem{kolter} L. Chai, H. Vlamakis and R. Kolter, Extracellular signal regulation of cell differentiation in biofilms, MRS Bulletin  36, 374-379, 2011.
\bibitem{daviesquorum} D.G. Davies, M.R. Parsek, J.P. Pearson, B.H. Iglewski,
J.W. Costerton, E.P. Greenberg, The involvement of cell-to-cell signals in the development of a bacterial biofilm, Science 280, 295-298, 1998.
\bibitem{daviesresistance} D. Davies, Understanding biofilm resistance to antibacterial agents, Nature Reviews 2, 114-122, 2003.
\bibitem{prebacillus} D.R. Espeso, A. Carpio, B. Einarsson, Differential growth
of wrinkled biofilms, Phys. Rev E 91, 022710, 2015.
\bibitem{scirep}  D.R. Espeso, A. Carpio, E. Martinez-Garcia, V. de Lorenzo,
Stenosis triggers spread of helical Pseudomonas biofilms in cylindrical flow systems,
Scientific Reports 6, 27170, 2016.
\bibitem{allen} M. A. A. Grant, B.Waclaw, R. J. Allen, and P. Cicuta, 
The role of mechanical forces in the planar-to-bulk transition in growing 
Escherichia coli microcolonies, J. R. Soc. Interface 11, 20140400, 2014.
\bibitem{halantolerance} B. Halan, A. Schmid,  K. Buehler,
Real-time solvent tolerance analysis of {\em Pseudomonas sp.} strain VLB120C catalytic biofilms, Appl. Env. Microbiol. 77, 1563-1571, 2011.
\bibitem{birnirmyxo} M. Hendrata, B. Birnir, Dynamic energy budget driven fruiting body formation in myxobacteria, Physical Review E 81, 061902, 2010.
\bibitem{hentzerinhibitquorum} M. Hentzer, K. Riedel, T.B. Rasmussen,
A. Heydorn, J.B. Andersen, M.R. Parsek, S.A. Rice, L. Eberl, S. Molin,  N.
H\o iby, S. Kjelleberg, M. Givskov, Inhibition of quorum sensing in {\em Pseudomonas aeruginosa} biofilm bacteria by a halogenated furanone 
compound, Microbiology 148, 87-102, 2002.
\bibitem{hoibyresistance} N. H\o iby, T. Bjarnsholt, M. Givskov, S. Molin, O. Ciofu, Antibiotic resistance of bacterial biofilms, International Journal of antimicrobial agents, Review, 322-332, 2010.
\bibitem{ishidadata} 
H. Ishida, Y. Ishida, Y. Kurosaka, T. Otani, K. Sato, H. Kobayashi,
In vivo and in vitro activities of levofloxacin against biofilm producing Pseudomonas aeruginosa, 
Antimicrob. Agents Chemother. 42, 1641-1645, 1998.
\bibitem{isidoritoxicity}  M. Isidori, M. Lavorgna, A. Nardelli, L. Pascarella, A. Parrella, Toxic and genotoxic evaluation of six antibiotics on non-target organisms, Science of the Total Environment 346, 87-98, 2005.
\bibitem{jaramayanantibioticaction} R. Jaramayan, Antibiotic resistance: an overview of mechanisms and a paradigm shift, Current Science 96, 1475-1484, 2009.
\bibitem{kimdoses} A. Kim, C.A. Sutherland, J.L. Kuti, D.P. Nicolau, Optimal dosing of piperacillin-tazobactam for the treatment of {\it Pseudomonas aeruginosa} infections: prolonged or continuous Infusion? Pharmacotherapy 27, 1490-1497, 2007.
\bibitem{klanjscekplos}  T. Klanjscek, R.M. Nisbet, J.H. Priester, P.A. Holden, Modeling physiological processes that relate toxicant exposure and bacterial population dynamics, PLOS One 7, e26955, 2012.
\bibitem{klanjscekcadmium} T. Klanjscek, R.M. Nisbet, J.H. Priester, P.A. Holden, Dynamic energy budget approach to modeling mechanisms of CdSe quantum dot toxicity, Ecotoxicology 22, 319-330, 2013.
\bibitem{kooijmanbook} S.A.L.M. Kooijman, Dynamic energy budget theory for metabolic organization, Cambridge University Press, 2008
\bibitem{ibmfluid} L. A. Lardon, B. V. Merkey, S. Martins, A. D\"otsch,
C. Picioreanu, J. U. Kreft, B. F. Smets, 
iDynoMiCS: next-generation individual-based modelling of biofilms,
Environ. Microbiol. 13, 241624-24, 2011.
\bibitem{kapellos} C. S. Laspidou and B. E. Rittmann, 
Modeling the development of biofilm density including active bacteria, inert biomass, and extracellular polymeric substances,
Water Res. 38, 3349-3361, 2004.
\bibitem{liinfluxeffluxpa} XZ Li, D Ma, DM Livermore, H Nikaido, Role of efflux pump(s) in intrinsic resistance of {\it Pseudomonas aeruginosa}: active efflux as a contributing factor to beta-lactam resistance, Antimicrob. Agents Chemother 38, 1742-1752, 1994.
\bibitem{lomovskayainhibitefflux}  O. Lomovskaya, M.S. Warren, A. Lee, J. Galazzo, R. Fronko, M. Lee, J. Balis, D. Cho, S. Chamberland, T. Renau, R. Leger, S. Hecker, W. Watkins, K. Hoshino, H. Ishida, V.J. Lee,
Identification and characterization of inhibitors of multidrug resistance efflux pumps in {\em Pseudomonas aeruginosa}: novel agents for combination therapy, Antimicrob. Agents Chemother. 45, 105-116, 2001.
\bibitem{mahamoudinhibitefflux} A. Mahamoud, J. Chevalier, S. Alibert-Franco, W.V. Kern, J.M. Page, Antibiotic efflux pumps in Gram-negative bacteria: the inhibitor response strategy, J. Antimicrob. Chemother.  59, 1223-1229, 2007.
\bibitem{mandsbergoxidative} L.F. Mandsberg, O. Ciofu, N. Kirby, L.E. Christiansen, H.E. Poulsen, N. H\o iby, Antibiotic resistance in {\em Pseudomonas aeruginosa} strains with increased mutation frequency due to inactivation of the DNA oxidative repair system, Antimicrob. Agents Chemother. 53, 2483-2491, 2009.
\bibitem{matrix} K.C. Marshall, Biofilms: an overview of bacterial adhesion, activity, and control at surfaces, ASM News 58, 202-207, 1992.
\bibitem{naganoinfluxeffluxec} K. Nagano, H. Nikaido, Kinetic behavior of the major multidrug efflux pump AcrB of Escherichia coli, Proc. Nat. Acad. Sc. 106, 5854-5858, 2009.
\bibitem{surveillance} National Nosocomial Infections Surveillance System. National nosocomial infections surveillance (NNIS) system report, data summary from January 1992 through June 2004, issued October 2004. Am J Infect Control 32, 470-485, 2004.
\bibitem{nicholsmodel} W.W. Nichols, M.J. Evans, M.P. Slack, H.L. Walmsley, The penetration of antibiotics into aggregates of mucoid and non-mucoid {\em Pseudomonas aeruginosa}. J Gen. Microbiol. 135, 1291-1303, 1989.
\bibitem{robinsonepsrates} J.A. Robinson, M.G. Trulear, W.G. Characklis, Cellular reproduction and extracellular polymer formation by {\it Pseudomonas aeruginosa} in continuous culture, Biotech. Bioeng. XXVI, 1409-1417, 1984.
\bibitem{stewartmodel} P.S. Stewart, Biofilm accumulation model that predicts antibiotic resistance of {\em Pseudomonas aeruginosa} biofilms, Antimicrob. Agents Chemother. 38, 1052-1057, 1994.
\bibitem{stewartresistance3} P.S. Stewart, J.W. Costerton, Antibiotic resistance of bacteria in biofilms, Lancet  358, 135-138, 2001.
\bibitem{stewartresistance4}  P.S. Stewart, Mechanisms of antibiotic resistance in bacterial biofilms, Int. J. Med. Microbiol. 292, 107-113, 2002.
\bibitem{stoneburden} P.W. Stone, Economic burden of healthcare-associated infections: An American perspective, Expert. Rev. Pharmacoeconomics Outcomes Res.  9, 417-422, 2009.
\bibitem{ibmsolid} T. Storck, C. Picioreanu, B. Virdis, and D. J. Batstone, 
Variable cell morphology approach for individual-based modeling of microbial communities, Biophys. J. 106, 2037-2048, 2014.
\bibitem{vickerydevice} K. Vickery, H. Hu, A.S. Jacombs, D.A. Bradshaw, A.K. Deva, A review of bacterial biofilms and their role in device-associated infection, Healthcare Infection  18, 61-66, 2013.
\bibitem{walterstolerance} M.C. Walters, F. Roe, A. Bugnicourt, M.J. Franklin, P.S. Stewart, Contributions of antibiotic penetration, oxygen limitation, and low metabolic activity to tolerance of {\em Pseudomonas aeruginosa} biofilms to ciprofloxacin and tobramycin, Antimicrob. Agents Chemother. 47, 317-323, 2003.
\bibitem{werneroxygen} E. Werner, F. Roe, A. Bugnicourt, M.J. Franklin, A. Heydorn, S. Molin,  B. Pitts, P.S. Stewart, Stratified growth in {\em Pseudomonas Aeruginosa} biofilms, Appl. Environ. Microbiol. 70, 6188-6196, 2004.

\bibitem{bib1}
Conant GC, Wolfe KH.
\newblock {{T}urning a hobby into a job: how duplicated genes find new
  functions}.
\newblock Nat Rev Genet. 2008 Dec;9(12):938--950.

\bibitem{bib2}
Ohno S.
\newblock Evolution by gene duplication.
\newblock London: George Alien \& Unwin Ltd. Berlin, Heidelberg and New York:
  Springer-Verlag.; 1970.

\bibitem{bib3}
Magwire MM, Bayer F, Webster CL, Cao C, Jiggins FM.
\newblock {{S}uccessive increases in the resistance of {D}rosophila to viral
  infection through a transposon insertion followed by a {D}uplication}.
\newblock PLoS Genet. 2011 Oct;7(10):e1002337.
\end{thebibliography}
\end{document}